\documentclass[preprint,superscriptaddress,footinbib,showkeys,showpacs,tightenlines]{revtex4} 
\usepackage{graphicx}
\usepackage{epstopdf}
\usepackage{dcolumn}
\newcommand{\be}{\begin{equation}}
\newcommand{\ee}{\end{equation}}
\newcommand{\bee}{\begin{eqnarray}}
\newcommand{\eee}{\end{eqnarray}}
\newcommand{\eq}{\end{quote}}
\newcommand{\nn}{\nonumber}
\newcommand{\eps}{\epsilon}
\newcommand{\Slash}[1]{\ooalign{\hfil/\hfil\crcr$#1$}}

\def\gsim{\displaystyle\mathop{>}_{\sim}}
\def\lsim{\displaystyle\mathop{<}_{\sim}}
\begin{document}      
\preprint{PNU-NTG-07/2005}
\title{Pentaquark $\Theta^+$ production via $\gamma N\to
\bar{K}^*\Theta^+(3/2^{\pm})$}  
\author{Seung-Il Nam}
\email{sinam@pusan.ac.kr}
\affiliation{Department of
Physics and Nuclear Physics \& Radiation Technology Institute (NuRI),
Pusan National University, Busan 609-735, Korea} 
\author{Atsushi Hosaka}
\email{hosaka@rcnp.osaka-u.ac.jp}
\affiliation{Research Center for Nuclear Physics (RCNP), Ibaraki, Osaka
567-0047, Japan}
\author{Hyun-Chul Kim}
\email{hchkim@pusan.ac.kr}
\affiliation{Department of
Physics and Nuclear Physics \& Radiation Technology Institute (NuRI),
Pusan National University, Busan 609-735, Korea} 
\date{May 2006}
\begin{abstract}
We study the photoproduction of the exotic pentaquark $\Theta^+$
baryon with the vector kaon, assuming that the quantum numbers of the
$\Theta^+$ to be $J^P=3/2^{\pm}$ and  $J^P=1/2^+$.  Scalar meson
$\kappa(800)$-exchange is also taken into account.  In contrast with
the $\gamma N\to \bar{K}\Theta^+(3/2^{\pm})$ process, the large suppression
from the proton target is not observed in the total cross
sections.  We also suggest a method
to determine which meson exchange is the most dominant by analyzing
the polarizations of incident photon and outgoing $K^*$.  
We find that  
$\kappa$-exchange turns out to be prominent when the
polarizations of the photon and $K^*$ are aligned to be parallel,
whereas $K$-exchange does when they are perpendicular to each other.    
\end{abstract}
\maketitle
\section{Introduction}
Since Diakonov {\it et al.} predicted the mass and width of the
pentaquark baryon $\Theta^+$~\cite{Diakonov:1997mm}, there has been a
great deal of works to clarify its existence and properties.  Although
various experiments have reported the existence of $\Theta^+$ after
the first observation by the LEPS collaboration~\cite{Nakano:2003qx},
the situation is not yet settled down primarily due to the relatively
low statistics of the low-energy experiments.  Furthermore, in almost
all high-energy experiments, the $\Theta^+$ has not been seen (see,
for example, a recent review~\cite{Hicks:2005gp,Schumacher:2005wu,
Hicks:2005jf} for the compilation of the experimental results). 

Recently, the CLAS collaboration has reported null results for finding
the $\Theta^+$ in the reactions 
$\gamma p \to \bar K^0 K^+ n$~\cite{Battaglieri:2005er}, 
$\gamma d \to \bar pK^- K^+ n$~\cite{McKinnon:2006zv}
and 
$\gamma d \to \bar \Lambda n K^+$~\cite{Niccolai:2006td}.  
The upper limits of the cross sections of producing 
$\Theta^+$ were estimetaed to be, for instance, 
$\sigma(\gamma p \to \bar K^0 \Theta^+) \sim 0.8$ nb, 
$\sigma(\gamma n \to \bar K^- \Theta^+) \sim 3$ nb.  
Though these experiments had high statistics, their results do not yet 
lead to the absence of $\Theta^+$ immediately, 
because the updatd positive evidences also seem rather convincing.   
In the LEPS, they observe a peak for the $\Theta^+$ in the reaction
$\gamma d \to \bar \Lambda(1520) n K^+$~\cite{nakano_jlab} when 
the $\Lambda(1520)$ is detected in the forward angle region.  
DIANA reported further evidence in 
the reaction $K^+ n\to K^0 p$ on a neutron bound in the Xenon
nucleus~\cite{Barmin:2006we}.
The statistical significance of the
DIANA measurement is $4.3\sim 7.3\,\sigma$. 
Moreover, 
KEK-PS E522 experiment has reported a measurement of
the $\Theta^+$ via the reaction 
$\pi^- p\to K^- X$~\cite{Miwa:2006if}, 
although the statistical siginificance is not large enough.  

Experimentally, the two similar experiments from CLAS and LEPS
are not in contradiction, since they measure different regions; 
CLAS detects final particles in the region where the scattering 
angle is not small, while the LEPS observes the forward angle region, 
and their measuring regions have little overlap.  
 

Theoretically, it was suggested that 
the production rate of 
the $\Theta^+$ from the proton target is considerably 
suppressed as compared to the case of the neutron target, 
if the spin of $\Theta^+$ is 3/2~\cite{Nam:2005jz}.  
Furthermore, in this case, the cross section of the neutron target 
which is larger than the proton case is strongly forward peaking.  
These may explain the different observations of the CLAS and LEPS.  
Interestingly, a similar suppression is found in the
$\Lambda(1520)$-photoproduction~\cite{Nam:2005uq}, though in this case  
the suppression takes place for the neutron target.
Therefore, it should be fare to say that the existance of 
the $\Theta^+$ is not yet excluded.  


Motivated by the previous work~\cite{Nam:2005uq}, we continue to
investigate the $\Theta^+$-photoproduction with the vector kaon $K^*$,
based on the effective Lagrangian approach with phenomenological form
factors.  Here, we consider the cases with $J^P=3/2^{\pm}$ and
$J^P=1/2^+$ for the $\Theta^+$ baryon. Scalar meson
$\kappa(800,0^+)$-exchange is also taken into account, in addition to
pseudoscalar $K$- and vector $K^*$-exchanges. We note that
$\kappa$-exchange in the $t$-channel does not appear in the $\gamma
N\to \bar{K}\Theta^+$ reaction process because the $\gamma\kappa K$
coupling is not allowed~\cite{Nam:2005jz}, whereas $\kappa$-exchange
is possible in the present reaction process according to the existence of
the $\gamma\kappa K^*$ coupling. The role of $\kappa$ may be
interesting if it is dominated by a tetraquark component which has
been suggested to have a strong coupling to exotic
baryons~\cite{Roy:2003hk}.      

One of the interesting features of the present reaction process is that 
there are two polarizations in the initial and final states: the
polarizations of the incident photon and the outgoing $K^*$. By making
a proper combination of these two polarizations, one can determine
which meson exchange in the $t$-channel dominates the reaction
process. 
            
The outline of the present work is sketched as follows: In Section 2,
we define the effective Lagrangians for the $\gamma N\to
\bar{K^*}\Theta^+(3/2^{\pm})$ reaction and calculate the invariant 
amplitudes with phenomenological form factors.  The numerical
results are given and discussed for the $\Theta^+(3/2^{\pm})$ and
$\Theta^+(1/2^+)$ in Section 3. Section 4 is devoted to the discussion 
on reaction analysis via the photon and $K^*$ polarizations. We
summarize our results and draw 
conclusion in the final Section.  
\section{Formalism}
\begin{figure}[t]
\resizebox{12cm}{6cm}{\includegraphics{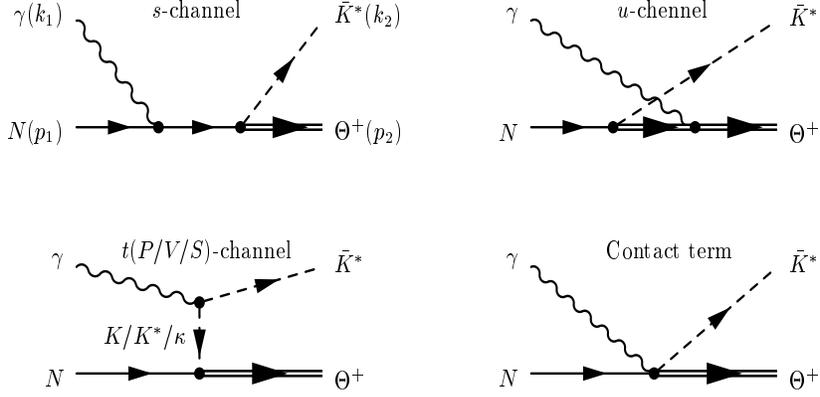}}
\caption{Born diagrams calculated in the effective Lagrangian
approach. $P/V/S$ in the $t$-channel stand for the pseudoscalar kaon,
vector kaon and scalar $\kappa$-exchanges, respectively.}    
\label{fig0}
\end{figure}
We investigate the reaction $\gamma N\to \bar{K}^*\Theta^+$ at the
tree level, i.e. in the Born approximation.  The relevant Feynman
diagrams are drawn in Fig.~\ref{fig0}, where we define  
the four momenta of the particles involved in the process. For
convenience, we will denote the spin $3/2$ and $1/2$ $\Theta^+$ with the
subscripts $3$ and $1$, respectively.  

The effective Lagrangians pertinent to the present work are given as
follows. { First, we consider the vertices of photon-meson-meson couplings:}

\bee
\mathcal{L}_{\gamma
  KK^{*}}&=&g_{\gamma
  KK^{*}}\epsilon_{\mu\nu\sigma\rho}(\partial^{\mu}A^{\nu})
(\partial^{\sigma}K)K^{*\rho}\,+{\rm 
h.c.},\\
\mathcal{L}_{\gamma
  K^*K^*}&=&ie[K^{*\dagger}_{\nu}(\partial_{\mu}K^{*\nu})-
K^{*}_{\nu}(\partial_{\mu}K^{*\dagger\nu})]A^{\mu},
\eee
where $K$, $K^*$ and $A^{\mu}$ denote the pseudoscalar kaon,  
vector kaon and photon fields, respectively. We employ the effective
Lagrangian taken from
Refs.~\cite{Liu:2003rh,Liu:2003zi,Oh:2003kw,Janssen:2001wk}. Note
that, in 
order to maintain gauge invariance of the reaction amplitudes, we
introduce a vector-meson exchange model using the $\gamma K^*K^*$  
vertex as shown in Eq.~(2), which was suggested by
Refs.~\cite{Clark:1970xr,Clark:1977fy}. This vertex represents three
vector particle coupling which manifests the nature of the non-Abelian
gauge fields.

The baryon electromagnetic couplings for the nucleon and the spin $3/2$
and $1/2$ $\Theta^+$ are defined as follows:

\bee
\mathcal{L}_{\gamma NN}
&=&-e\bar{N}\left[\Slash{A}+\frac{\kappa_N}{4M_{N}}
\sigma_{\mu\nu}F^{\mu\nu}\right]N\,+{\rm
h.c.},\\
\mathcal{L}_{\gamma
\Theta_1\Theta_1}&=&-e\bar{\Theta}_1\left[\Slash{A}+
\frac{\kappa_{\Theta}}{4M_{\Theta}}\sigma_{\mu\nu}
F^{\mu\nu}\right]\Theta_1\,+{\rm 
h.c.},\\
\mathcal{L}_{\gamma
\Theta_3\Theta_3}&=&-e\bar{\Theta}^{\mu}_3g_{\mu\nu}\left[\Slash{A}+ 
\frac{\kappa_{\Theta}}{4M_{\Theta}}\sigma_{\sigma\rho}
F^{\sigma\rho}\right]\Theta^{\nu}_3\,+{\rm 
h.c.},
\eee
where $N$, $\Theta^{\mu}_3$ and $\Theta_1$ stand for the nucleon, the spin
$3/2$ 
Rarita-Schwinger (RS) $\Theta^+$~\cite{Read:ye} and spin $1/2$
$\Theta^+$, respectively. The same structures of the Lagrangians  are
used for the nucleon and  spin $1/2$ $\Theta^+$ (Eqs.~(3,4)). Following
Ref.~\cite{gourdin}, we 
construct the effective Lagrangian for the electromagnetic coupling of
the spin $3/2$ $\Theta^+$ in Eq.~(5).  Here, being different from
Ref.~\cite{gourdin}, since the electric
quadrapole ($E2$) and magnetic octupole ($M3$) form factors are
expected to be  small, compared to the charge and magnetic 
dipole form factors of spin $3/2$ baryons, we only consider the
$E0$ and $M1$ electromagnetic interaction. Concerning other
possible structures of the electromagnetic 
couplings, it
is worth mentioning that, as indicated in 
Ref.~\cite{Read:ye}, 
the electromagnetic coupling for the spin $3/2$ $\Theta^+$ can be
reconstructed equivalently with the terms such as 
$\bar{\Theta}^{\mu}F_{\mu\nu}\Theta^{\nu}$ and others.

The $K(K^*)N\Theta$ vertices for the spin $3/2$ and 1/2 $\Theta^+$ baryons
are defined as follows~\cite{Nam:2005uq,Machleidt:1987hj}. 

\bee
\mathcal{L}_{KN\Theta_3}&=&\frac{g_{KN\Theta_3}}{M_{K}}\bar{\Theta}^{\mu}_3
\partial_{\mu}K{\Gamma_5}N\,+{\rm 
h.c.},\\
\mathcal{L}_{KN\Theta_1}&=&ig_{KN\Theta_1}
\bar{\Theta}_1\Gamma_5\gamma_5KN+{\rm h.c.},\\
\mathcal{L}_{K^{*}N\Theta_3}&=&-\frac{ig_{K^{*}N\Theta_3}}
{M_{K^*}}\bar{\Theta}_{3,\mu}\gamma_{\nu}F^{\mu\nu}_{K^*}\Gamma_5\gamma_5N+{\rm
  h.c.},\\
\mathcal{L}_{K^*N\Theta_1}&=&g^V_{K^*N\Theta_1}
\bar{\Theta}_1\gamma_{\mu}\Gamma_5K^{*\mu}N-\frac{g^T_{K^*N\Theta_1}}{2(M_{\Theta}+M_N)}\bar{\Theta}_1\Gamma_5\sigma_{\mu\nu}F^{\mu\nu}_{K^*}N+{\rm 
h.c.}, 
\eee
where $\Gamma_5$ denotes ${\bf 1}_{4\times4}$ in the 
positive-parity and $\gamma_5$ for the negative-parity  
$\Theta^+$, respectively, for both cases of the spin $3/2$ and
spin $1/2$. $F^{\mu\nu}_{K^*}$ stands for
$\partial^{\mu}K^{*\nu}-\partial^{\nu}K^{*\mu}$. As
for the spin $1/2$ $\Theta^+$, we consider only the pseudoscalar
coupling scheme for the $KN\Theta$ vertex due to the approximate
equivalence between the pseudoscalar and psedovector
schemes~\cite{Nam:2003uf}. On the 
contrary, only pseudovector (derivative) coupling is possible for the
case of the 
spin $3/2$ due to the constraint
$\gamma_{\mu}\Theta^{\mu}=0$. Concerning the $K^*N\Theta$ vertex of
Eqs.~(8,9), we 
consider the Lagrangian structures which are necessary
minimally for maintaining the gauge invariance when we construct
reaction amplitudes. Note that, as for the
spin $1/2$ case, we have the vector and tensor terms in
the Lagrangian of Eq.~(9). Here, we use the value of
$g^T_{K^*N\Theta_1}=|g^V_{K^*N\Theta_1}|$ as a trial since no
experimental data are available now. However, the strength of
$g^T_{K^*N\Theta}$ can be estimated from the recent 
calculations of the transition magnetic moment of $\gamma
N_8N^*_{\bar{10}}$, where $\kappa_{\gamma N_8N^*_{\bar{10}}}$ was found 
to be $0\sim0.5$~\cite{Choi:2005ki,Kim:2005gz,Azimov:2005jj}. Here,
$N^*_{\bar{10}}$  
is a nucleon partner of the antidecuplet pentaquark. Assuming the
vector dominance and flavor SU(3) symmetry, we expect that the ratio
$|g^T_{K^*N\Theta}/g^V_{K^*N\Theta}|$ is less than unity. Thus, our
choice of $g^T_{K^*N\Theta_1}=|g^V_{K^*N\Theta_1}|$ can be almost its upper
bound. 

Finally, we introduce the photon coupling in the $K^*N\Theta$
vertex by minimal
substitution, $\partial_{\mu}\to\partial_{\mu}+i\hat{Q}A_{\mu}$ where
$\hat{Q}$ is the charge matrix acting upon the matter fields.

\bee
\mathcal{L}_{\gamma K^{*}N\Theta_3}&=&\frac{eg_{K^{*}N\Theta_3}}
{M_{K^*}}\bar{\Theta}^{\mu}_3\gamma^{\nu}[A_{\mu}
K^{*}_{\nu}-A_{\nu}K^{*}_{\mu}]\Gamma_5\gamma_5N+{\rm 
h.c.},\\
\mathcal{L}_{\gamma
K^{*}N\Theta_1}&=&-\frac{ieg^T_{K^*N\Theta_1}}{2(M_{\Theta}+M_N)}\bar{\Theta}_1\Gamma_5\sigma_{\mu\nu}(A^{\mu}K^{*\nu}-A^{\nu}K^{*\mu})N+{\rm
h.c.}, 
\eee
These interaction vertices are related to the Feynman diagram of the
contact term shown in Fig.~\ref{fig0}. We note that the same
interactions of Eqs.~(10,11) are obtained from the non-Abelian terms
of the covariant field tensor
$\partial_{\mu}V_{\nu}-\partial_{\nu}V_{\mu}-i[V_{\mu},V_{\nu}]$ where
$V_{\mu}$ is an SU(3) vector meson field, and using the vector dominance.

\begin{table}[t]
\begin{tabular}{|c|c|cc|cccccc|}
\hline
&$\kappa_N$&&$g_{\gamma KK^*}$&&$g_{KN\Theta_3}$&$g^V_{K^*N\Theta_3}$
&$g_{KN\Theta_1}$&$g^V_{K^*N\Theta_1}$&$g^T_{K^*N\Theta_1}$\\
\hline
$n$&$-$1.91&Neutral&0.388/GeV&$\pi(\Theta)=+1$&0.53&0.91=0.53$\sqrt{3}$&1&$\sqrt{3}$&$\sqrt{3}$\\   
$p$&1.79&Charged&0.254/GeV&$\pi(\Theta)=-1$&4.22&2.0&$-$&$-$&$-$\\
\hline
\end{tabular}
\caption{Parameters of the couplings used in the numerical calculations}
\label{table1}
\end{table}
In Table.~\ref{table1}, we
list the parameters (elecgromagnetic and strong couplings) which are used
for numerical calculation. The nucleon magnetic moments $\kappa_N$ and
the $\gamma K^*K$ coupling 
constants are taken from experiments~\cite{Eidelman:2004wy}. For
$g_{KN\Theta}$, we assume 
$\Gamma_{\Theta\to KN}=1$ MeV and $M_{\Theta}=1540$ MeV for both
spin $3/2$ and 1/2~\cite{Eidelman:2004wy}. For $g^V_{K^*N\Theta}$, we
assume the estimation in 
the quark model $g^V_{K^*N\Theta}=\sqrt{3}g_{KN\Theta}$ for the
positive parity $\Theta^+$~\cite{Close:2004tp} whereas we used the results of
Ref.~\cite{Hosaka:2004bn} for $\Theta(3/2^-)$. As for the value of the
anomalous magnetic moment 
of $\Theta^+$, we 
set it to be unity for both spins as a trial.  We will
show later that the dependence on $\kappa_{\Theta}$ is negligible,
since the $u$-channel contributions turn out to be very small. Since we 
verified that the sign of $g^V_{K^*N\Theta}$ does not influence much
on the results as shown in the previous work~\cite{Nam:2005jz}, we will
only consider plus sign. The case of $\Theta(1/2^-)$ is not studied since
we verified that it behaves very similar to that of $\Theta(1/2^+)$
except for the only obvious difference in the order of magnitudes
being smaller by factor about ten~\cite{Nam:2003uf}. We note that in
the present work, we 
do not consider nucleon resonance ($N^*$) contributions. In other
words, we only 
take into account the minimally possible reaction diagrams as shown in
Fig.~\ref{fig0}.  

Thus,
the reaction amplitudes for spin $3/2$ ($\mathcal{M}_3$)  and 
1/2  ($\mathcal{M}_1$) can be written as follows. Furthermore, we have checked
that the amplitudes calculated 
from the Lagrangians satisfy the Ward-Takahashi identity with the form
factors.  

\bee
i\mathcal{M}_{3,s}&=&-\frac{ieg_{K^*N\Theta}}{M_{K^*}}
\bar{u}(p_2)[(k_{2}\cdot\epsilon_{\Theta})\Slash{\epsilon}_{K^*}
-(\epsilon_{\Theta}\cdot\epsilon_{K^*})\Slash{k}_{2}]\Gamma_5
\gamma_5\frac{(\Slash{p}_1+M_N)F_c+\Slash{k}_1F_s}{q^2_s-M^2_N}\Slash{k}_1u(p_1)\nn\\&-&\frac{ie\kappa_Ng_{K^*N\Theta}}{2M_NM_{K^*}}
\bar{u}(p_2)[(k_{2}\cdot\epsilon_{\Theta})\Slash{\epsilon}_{K^*}
-(\epsilon_{\Theta}\cdot\epsilon_{K^*})\Slash{k}_{2}]\Gamma_5
\gamma_5\frac{(\Slash{q}_s+M_N)F_s}{q^2_s-M^2_2}
\Slash{\epsilon}_{\gamma}\Slash{k}_1u(p_1),\nn\\
i\mathcal{M}_{3,u}&=&-\frac{ieg_{K^*N\Theta}}{M_{K^*}}\bar{u}(p_2)
\Slash{\epsilon}_{\gamma}\frac{(\Slash{p}_2+M_{\Theta})F_c+
\Slash{k}_1F_u}{q^2_u-M^2_{\Theta}}[(k_{2}\cdot\epsilon_{\Theta})\Slash{\epsilon}_{K^*}-(\epsilon_{\Theta}\cdot\epsilon_{K^*})\Slash{k}_2]\Gamma_5\gamma_5u(p_1)\nn\\&-&
\frac{ie\kappa_{\Theta}g_{K^*N\Theta}}{2M_{\Theta}M_{K^*}}\bar{u}(p_2)
\Slash{\epsilon}_{\gamma}\Slash{k}_1\frac{(\Slash{q}_u+M_{\Theta})F_u}{q^2_u-M^2_{\Theta}}[(k_{2}\cdot\epsilon_{\Theta})\Slash{\epsilon}_{K^*}-(\epsilon_{\Theta}\cdot\epsilon_{K^*})\Slash{k}_2]\Gamma_5\gamma_5u(p_1),\nn\\
i\mathcal{M}_{3,t(P)}&=&-\frac{g_{\gamma KK^*}g_{KN\Theta}}{M_K}
\frac{\bar{u}(p_2)\Gamma_5u(p_1)}{q^2_t-M^2_K}
\left[(\epsilon_{\Theta}\cdot q_t)\epsilon_{\mu\nu\sigma\rho}
k^{\mu}_1\epsilon^{\nu}_{\gamma}q^{\sigma}_t\epsilon^{\rho}_{K^*}\right]F_t,\nn\\
i\mathcal{M}_{3,t(V)}&=&-\frac{ieg_{K^*N\Theta}}{M_{K^*}}\bar{u}(p_2)
\frac{2\epsilon_{\gamma}\cdot k_2}{q^2_t-M^2_{k^*}}[(q_{t}\cdot
\epsilon_{\Theta})\Slash{\epsilon}_{K^*}-(\epsilon_{\Theta}\cdot
\epsilon_{K^*})\Slash{q}_t]\Gamma_5\gamma_5u(p_1)F_c,\nn\\
i\mathcal{M}_{3,c}&=&-\frac{ieg_{K^*N\Theta}}{M_{K^*}}
\bar{u}(p_2)[(\epsilon_{\gamma}\cdot\epsilon_{\Theta})
\Slash{\epsilon}_{K^*}-(\epsilon_{\Theta}\cdot\epsilon_{K^*})
\Slash{\epsilon}_{\gamma}]\Gamma_5\gamma_5u(p_1)F_c.\nn\\
\label{amplitudes3}
\eee

\bee
i\mathcal{M}_{1,s}&=&ieg^V_{K^*N\Theta_1}\bar{u}(p_2)
\Slash{\epsilon}_{K^*}\Gamma_5\frac{(\Slash{p}_1+M_N)F_c+
\Slash{k}_1F_c}{q^2_s-M^2_N}\Slash{\epsilon}_{\gamma}u(p_2)\nn\\&+&\frac{ie\kappa_Ng^V_{K^*N\Theta_1}}{2M_N}\bar{u}(p_2)
\Slash{\epsilon}_{K^*}\Gamma_5\frac{(\Slash{q}_s+M_N)F_s}{q^2_s-M^2_N}\Slash{k}_1\Slash{\epsilon}_{\gamma}u(p_2)\nn\\&+&\frac{ieg^T_{K^*N\Theta_1}}{2(M_{\Theta}+M_N)}\bar{u}(p_2)\Gamma_5{(\Slash{k}_2\Slash{\epsilon}_{K^*}-\Slash{\epsilon}_{K^*}\Slash{k}_2)}\frac{(\Slash{p}_1+M_N)F_c+\Slash{k}_1F_s}{q^2_s-M^2_N}\Slash{\epsilon}_{\gamma}u(p_2)\nn\\&-&
\frac{ie\kappa_Ng^T_{K^*N\Theta_1}}{4M_N(M_{\Theta}+M_N)}\bar{u}(p_2)\Gamma_5(\Slash{k}_2\Slash{\epsilon}_{K^*}-\Slash{\epsilon}_{K^*}\Slash{k}_2)\frac{\Slash{p}_1+\Slash{k}_1+M_N}{q^2_s-M^2_N}\Slash{\epsilon}_{\gamma}\Slash{k}_1u(p_2)F_s,\nn\\
i\mathcal{M}_{1,u}&=&ieg^V_{K^*N\Theta_1}\bar{u}(p_2)\Slash{\epsilon}_{\gamma}\frac{(\Slash{p}_s+M_{\Theta})F_c-\Slash{k}_1F_u}{q^2_u-M^2_{\Theta}}\Slash{\epsilon}_{K^*}\Gamma_5u(p_1)\nn\\&+&\frac{ie\kappa_{\Theta}g^T_{K^*N\Theta_1}}{4M_{\Theta}(M_{\Theta}+M_N)}\bar{u}(p_2)\Slash{k}_1\Slash{\epsilon}_{\gamma}\frac{(\Slash{q}_u+M_{\Theta})F_s}{q^2_u-M^2_{\Theta}}\Slash{\epsilon}_{K^*}\Gamma_5u(p_1)\nn\\&+&\frac{ieg^T_{K^*N\Theta_1}}{2(M_{\Theta}+M_N)}\bar{u}(p_2)\Slash{\epsilon}_{\gamma}\frac{(\Slash{p}_2+M_{\Theta})F_c-\Slash{k}_1F_u}{q^2_u-M^2_{\Theta}}\Gamma_5{(\Slash{k}_2\Slash{\epsilon}_{K^*}-\Slash{\epsilon}_{K^*}\Slash{k}_2)}u(p_2)\nn\\&-&
\frac{ie\kappa_{\Theta}g^T_{K^*N\Theta_1}}{4M_{\Theta}(M_{\Theta}+M_N)}\bar{u}(p_2)\Slash{k}_1\Slash{\epsilon}_{\gamma}\frac{\Slash{p}_2-\Slash{k}_1+M_{\Theta}}{q^2_u-M^2_{\Theta}}\Gamma_5(\Slash{k}_2\Slash{\epsilon}_{K^*}-\Slash{\epsilon}_{K^*}\Slash{k}_2)u(p_2)F_u,\nn\\
\mathcal{M}_{1,t(P)}&=&g_{KN\Theta_1}g_{\gamma
KK^*}\frac{\bar{u}(p_1)\Gamma_5\gamma_5
u(p_1)}{q^2_t-M^2_K}\epsilon_{\mu\nu\rho\sigma}k^{\mu}_1
\eps^{\nu}_{\gamma}\eps^{\rho}_{K^*}q^{\sigma}_tF_t\nn,\nn\\ 
\mathcal{M}_{1,t(V)}&=&-2ieg_{K^*N\Theta_1}\bar{u}(p_1)\frac{k_2\cdot\eps_{\gamma}
\Slash{\eps}_{K^*}\Gamma_5}{q^2_t-M^2_{K^*}}u(p_1)F_c\nn\\&+&\frac{ie\kappa_Ng^T_{K^*N\Theta_1}}{M_{\Theta}+M_N}\bar{u}(p_2)\Gamma_5(\Slash{q}_t\Slash{\epsilon}_{K^*}-\Slash{\epsilon}_{K^*}\Slash{q}_t)\frac{k_2\cdot\epsilon_{\gamma}}{q^2_t-M^2_{K^*}}F_cu(p_1),\nn\\
\mathcal{M}_{1,c}&=&\frac{ieg^T_{K^*N\Theta_1}}{2(M_{\Theta}+M_N)}\bar{u}(p_2)\Gamma_5(\Slash{\epsilon}_{\gamma}\Slash{\epsilon}_{K^*}-\Slash{\epsilon}_{K^*}\Slash{\epsilon}_{\gamma})u(p_1).
\label{amplitudes1}
\eee
The subscripts $s$, $u$, $t(P)$, $t(V)$, and $c$ of $\mathcal{M}$
indicate $s$-, 
$u$-, pseudoscalar $K$-exchange, vector $K^*$-exchange and the contact term,
respectively.  $q_s=p_1+k_1$, $q_t=k_1-k_2$ 
and $q_u=p_1-k_2$ are the momentum transfers for each kinematical  
channel.  The Mandelstam variables $s$, $t$, and $u$ are defined in a standard
way: $s=q^2_s$, $u=q^2_u$ and $t=q^2_t$.  For spin $3/2$ 
$\Theta^+$, we need to take into account 
$\mathcal{M}_{s,E,M}$, $\mathcal{M}_{u,E,M}$ and $\mathcal{M}_{t(P)}$ for the 
proton target and $\mathcal{M}_{s,M}$, $\mathcal{M}_{u,E,M}$,
$\mathcal{M}_{t(P)}$, $\mathcal{M}_{t(V)}$ and $\mathcal{M}_c$ for 
the neutron one, where $E$ and $M$ stand for the terms including electric 
(proportional to $e$) and magnetic (proportional to
$e\kappa_{N,\Theta}$) interactions.  $\epsilon_{\gamma}$ and
$\epsilon_{K^*}$ are the  
polarization vectors of the photon and the vector
kaon, respectively.  $\epsilon_{\Theta}$ is the spin-1 component of
the Rarita-Schwinger field for the $\Theta^+$~\cite{Nam:2005uq}. We
simplify the spin $3/2$ RS propagator by that of spin $1/2$ baryon. It was 
shown that this simplification worked qualitatively well in the low-energy
regions~\cite{Nam:2005uq}. The 
evaluation of the invariant amplitudes for the
spin $1/2$ is also performed similarly to that of spin $3/2$.

In the present work, we also take into
account scalar meson 
$\kappa(800,0^+)$-exchange in addition to $K$- and 
$K^*$-exchange.  The relevant effective Lagrangians are defined as
follows: 

\bee
\mathcal{L}_{\gamma\kappa K^*}&=&g_{\gamma\kappa
  K^*}F_{\mu\nu}F^{\mu\nu}_{K^*}\kappa,\nn\\ 
\mathcal{L}_{\kappa N\Theta_3}&=&\frac{g_{\kappa N\Theta_3}}
{M_{\kappa}}\bar{\Theta}^{\mu}_3(\partial_{\mu}\kappa)
\Gamma_5\gamma_5N,\nn\\
\mathcal{L}_{\kappa N\Theta_1}&=&ig_{\kappa
N\Theta_1}\bar{\Theta}_1\Gamma_5\kappa N, 
\label{kappa}
\eee
where $\kappa$ indicates the scalar meson field with its
physical mass $\sim 800$ MeV~\cite{Eidelman:2004wy}. Since there is no
information of the coupling constants 
$g_{\gamma\kappa K^*}$ and $g_{\kappa N\Theta_{3}}$, we will estimate
them for both the spin $3/2$ and 1/2 $\Theta^+$ as follow as a trial:

\bee
g_{\gamma\kappa K^*}=|g_{\gamma KK^*}|\,\,\, {\rm and}\,\,\,g_{\kappa N\Theta_{3}}=|g_{KN\Theta_{3}}|.\nn
\eee

We note that the signs of these coupling constants are unknown
and not estimated by flavor SU(3) symmetry. However, we
verified that the signs of these coupling constants do not make 
significant differences in the numerical results. Hence, we only
consider plus 
signs for the coupling constants. The reaction 
amplitudes for $\kappa$-exchange (t(S)) can be written as follows: 

\bee
i\mathcal{M}_{3,t(S)}&=&-\frac{2g_{\gamma\kappa K^*}g_{\kappa
    N\Theta_3}}{M_{\kappa}} \frac{\bar{u}(p_2)\Gamma_5\gamma_5u(p_1)}
{q^2_t-M^2_{\kappa}}[\eps_{\Theta}\cdot q_t][(k_1\cdot
k_2)(\eps_{\gamma}\cdot\eps_{K^*})-(\eps_{\gamma}\cdot k_2)(\eps_{K^*}
\cdot k_1)]F_{\kappa},\nn\\
i\mathcal{M}_{1,\kappa}&=&-2ig_{\gamma\kappa 
K^*}g_{\kappa N\Theta_1}\frac{\bar{u}(p_2)\Gamma_5u(p_1)}{q^2_t-M^2_{\kappa}}[(k_1\cdot
k_2)(\eps_{\gamma}\cdot\eps_{K^*})-(\eps_{\gamma}\cdot k_2)(\eps_{K^*}\cdot
k_1)].
\label{amplitudes}
\eee

As shown in Eqs.~(\ref{amplitudes3}), (\ref{amplitudes1}) and
(\ref{amplitudes}), we employ the four-dimensional form
factors~\cite{Nam:2005uq} defined as follows: 

\bee
F_{x}(q^2)&=&\frac{\Lambda^4}{\Lambda^4+(x-M^2_x)^2},\,\,\,
x=s,t,u,\nn\\ 
F_{c}&=&F_u+F_{t(V)}-F_uF_{t(V)}\,\,\,{\rm for\,\,neutron},\nn\\
F_{c}&=&F_s+F_u-F_sF_u\,\,\,{\rm for\,\,proton},
\label{formfactor}
\eee        
where $M_x$ is the mass of the interchanged particle in the $x$-channels. We
verified that the inclusion of the form factor maintains the gauge 
invariance. We make use of the
cutoff value $\Lambda=750$ MeV as in
Refs.~\cite{Nam:2005jz,Nam:2005uq}. 
\section{Numerical results}
We present in this section the numerical results of the total and
differential cross sections, asymmetries, and momentum transfer
$t$-dependences for the neutron and proton targets.  Here, the
asymmetry is defined as follows:   

\bee
{\rm Asymmetry}=\frac{\left(\frac{d\sigma}{d\Omega}\right)_{\perp}
-\left(\frac{d\sigma}{d\Omega}\right)_{\parallel}}{
\left(\frac{d\sigma}{d\Omega}\right)_{\perp}
+\left(\frac{d\sigma}{d\Omega}\right)_{\parallel}}.
\label{asym}
\eee
The notations $\parallel$ and $\perp$ in Eq.~(\ref{asym}) stand for 
the photon polarizations which are parallel and perpendicular to the
reaction plane, respectively.  

\begin{figure}[t]
\begin{tabular}{cc}
\includegraphics[width=7cm]{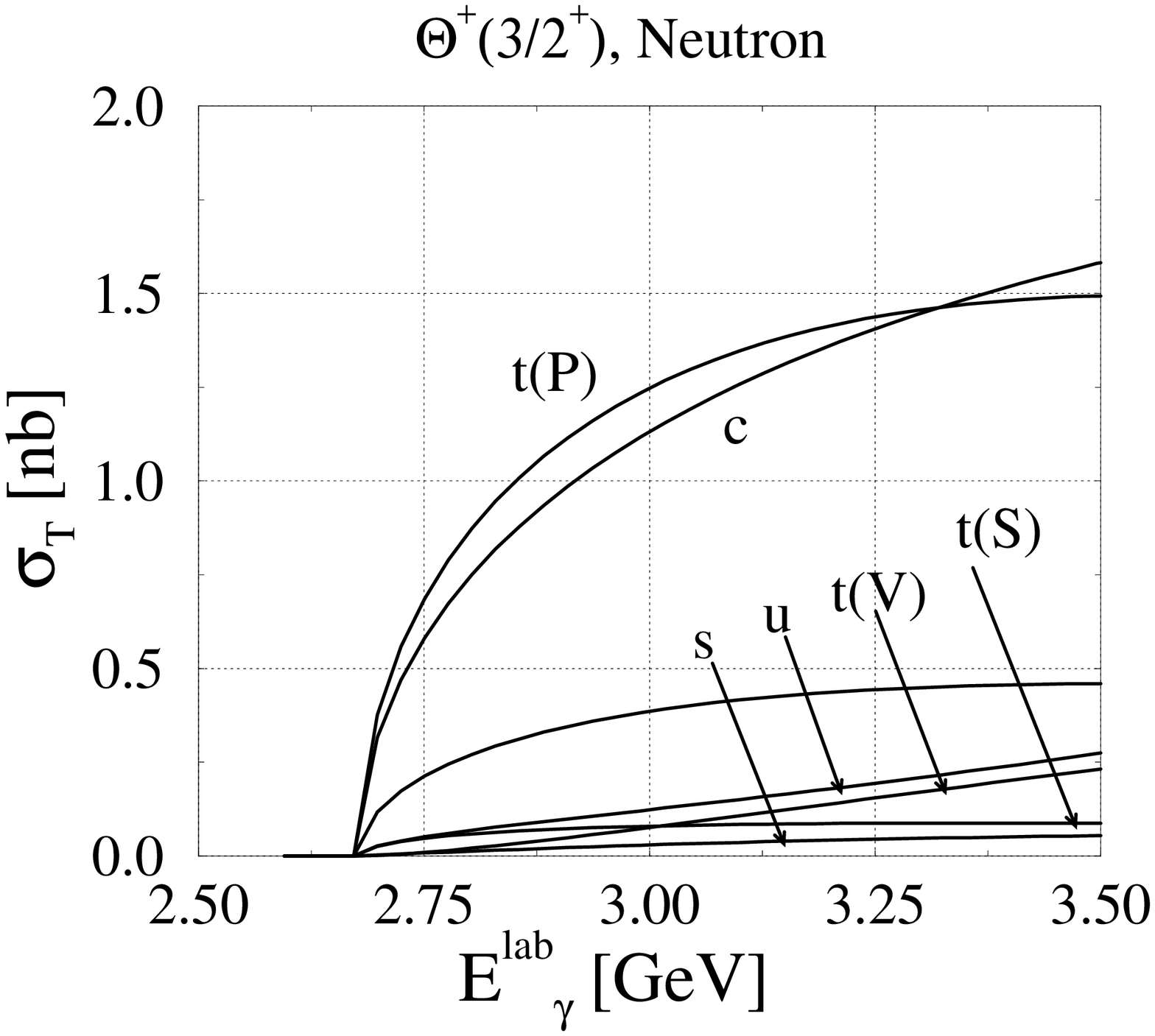}
\includegraphics[width=7cm]{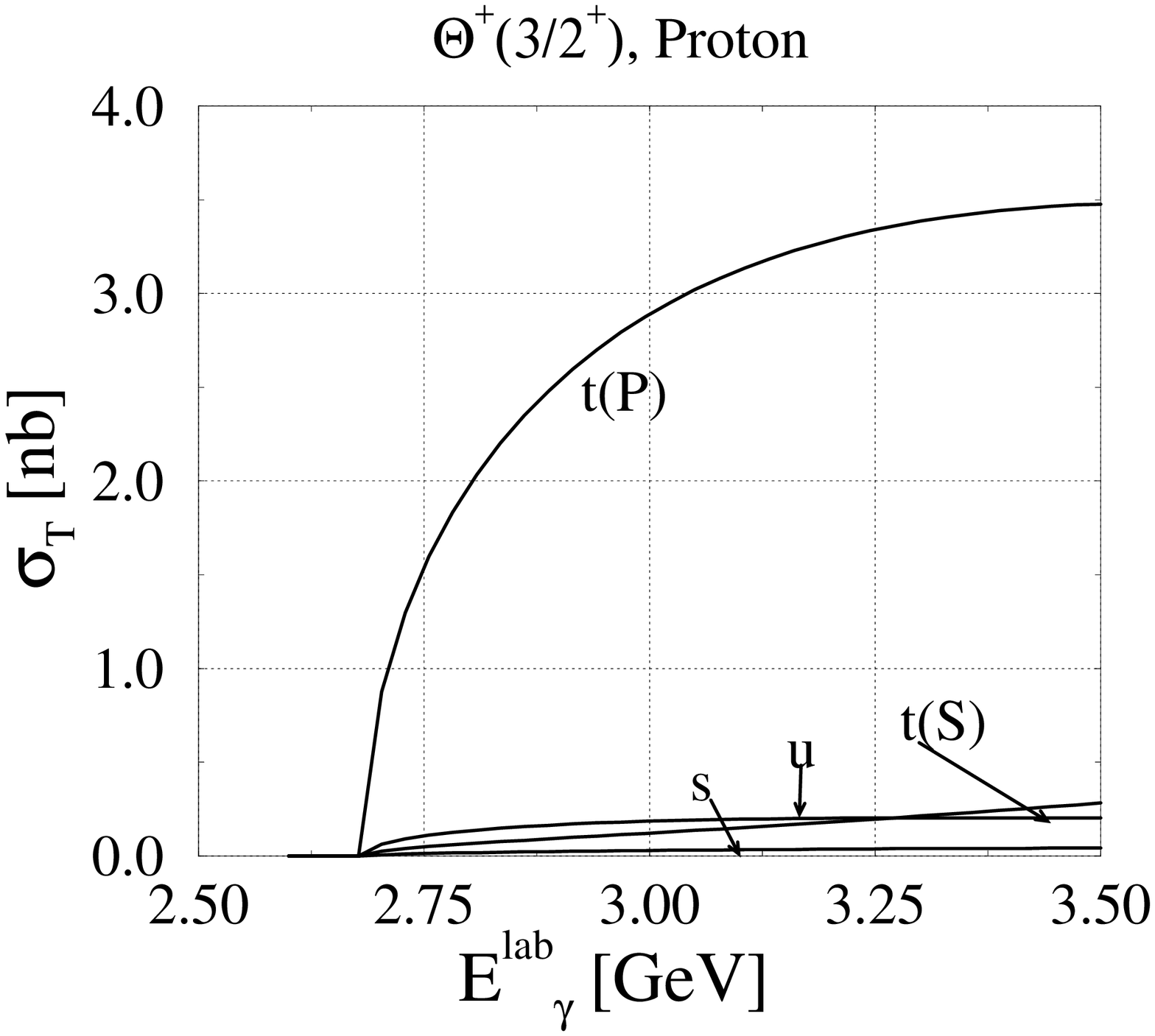}
\end{tabular}
\begin{tabular}{cc}
\includegraphics[width=7cm]{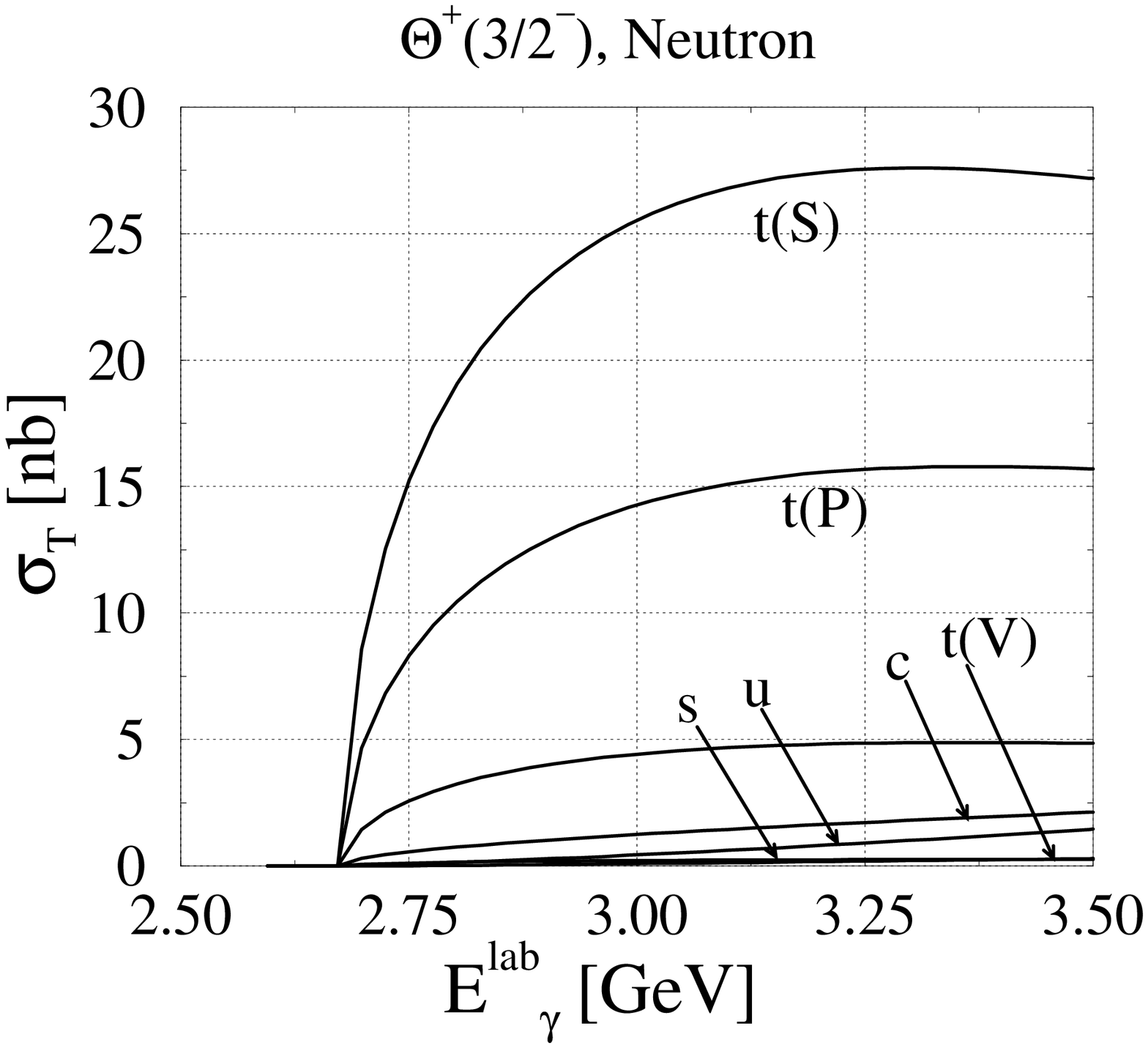}
\includegraphics[width=7cm]{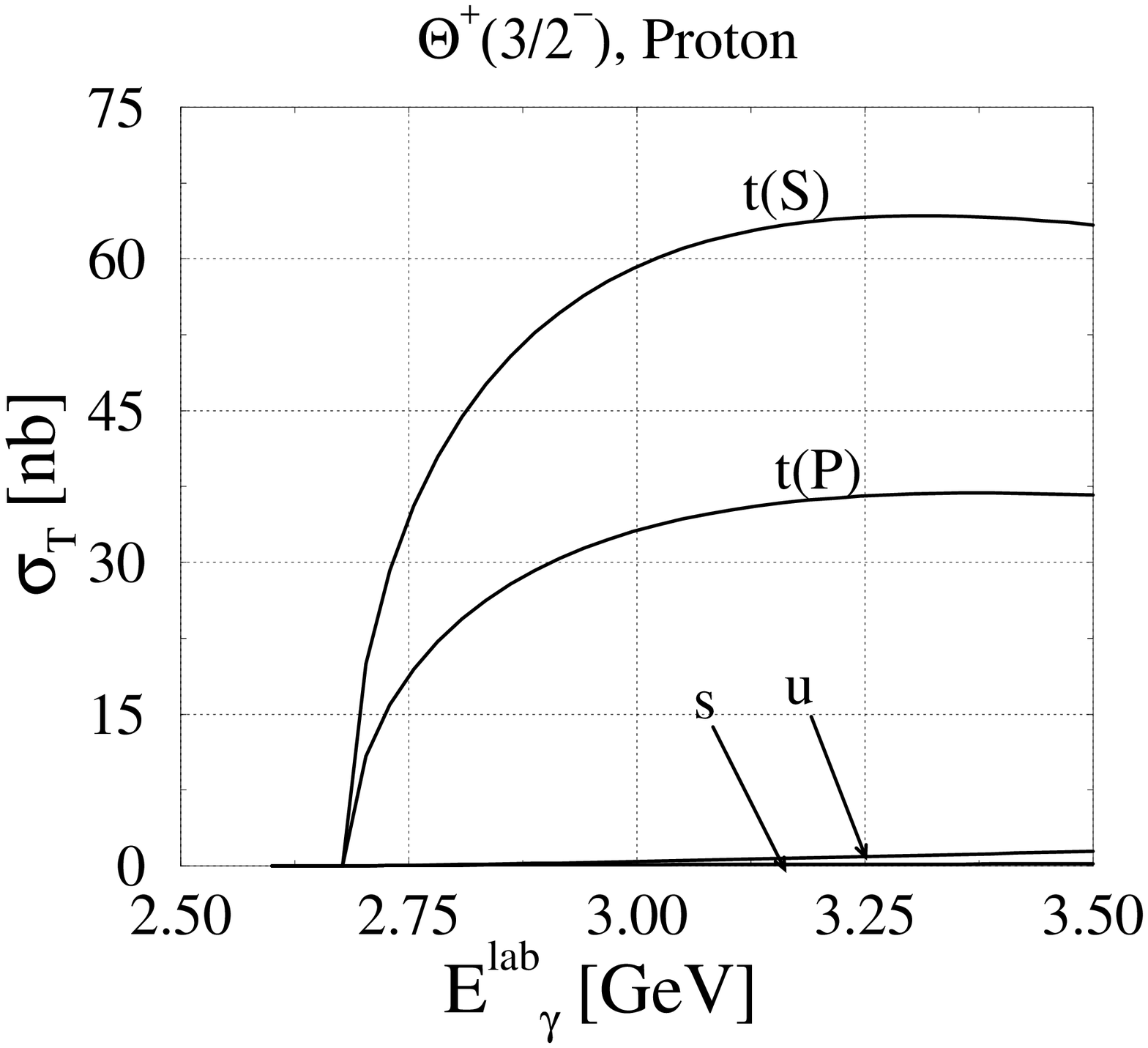}
\end{tabular}
\caption{Various contributions to the total cross sections from
different kinematical channels. The labels are defined by
s ($s$-channel), u ($u$-channel), t(P) (pseudoscalar kaon exchange in
$t$-channel), t(V) (vector kaon exchange in
$t$-channel), t(S) (scalar $\kappa$ exchange in
$t$-channel) and c (contact term). We 
show the four different cases, i.e. $\Theta^+(3/2^+)$ from the neutron
(upper-left) and 
proton (upper-right) targets, and $\Theta^+(3/2^-)$ from  the neutron
(lower-left) and proton (lower-right) ones.}       
\label{fig1}
\end{figure}

In Fig.~\ref{fig1}, we show various contributions to the total
cross sections for each kinematical channel separately as 
functions of photon energy in the laboratory frame ($E^{\rm
  lab}_{\gamma}$).  The upper two panels represent the results for
the $\Theta^+(3/2^+)$, where we see that the contact and
psuedoscalar $K$-exchange terms are main contributions for the neutron 
target, whereas the $K$-exchange term dominates the reaction for
the proton one.  Since the $\gamma K^*K$ 
coupling constants for the proton and neutron targets differ by
$g_{\gamma K^0\bar{K}^{*0}}/g_{\gamma K^+K^{*-}}\sim 1.5$, we obtain
the 
contribution of $K$-exchange to the total cross sections for the proton 
target about two times larger than the neutron one.  Being 
different from $\Theta^+(3/2^+)$, $\kappa$- and 
$K$-exchanges govern the reaction for the $\Theta^+(3/2^-)$ as
demonstrated in the lower two panels.  The total cross sections of 
$K$-exchange for $\Theta^+(3/2^-)$ becomes much larger than those of
$\Theta^+(3/2^+)$ due to the $d$-wave coupling for the $KN\Theta_3$
vertex.  The large  contribution of
$\kappa$-exchange can be understood by that we assumed larger coupling
constants $g_{\kappa N\Theta}$ and $g_{\kappa\gamma K^*}$ for
$\Theta^+(3/2^-)$ than those of $\Theta^+(3/2^+)$.  However, even if 
we ignore $\kappa$-exchange, the qualitative tendency
$\sigma_{3/2^+}<\sigma_{3/2^-}$ will not be altered, since
$K$-exchange is more dominant than the contributions from the
$\kappa$-exchange.  Moreover, though we can see about two or
three times
difference in magnitudes of the total cross sections between the
neutron and proton targets, the difference is much smaller than that
of the $\Lambda^*$-photoproduction associated with the pseudoscalar kaon as shown
in the previous work~\cite{Nam:2005jz}.

\begin{figure}[t]
\begin{tabular}{cc}
\includegraphics[width=7cm]{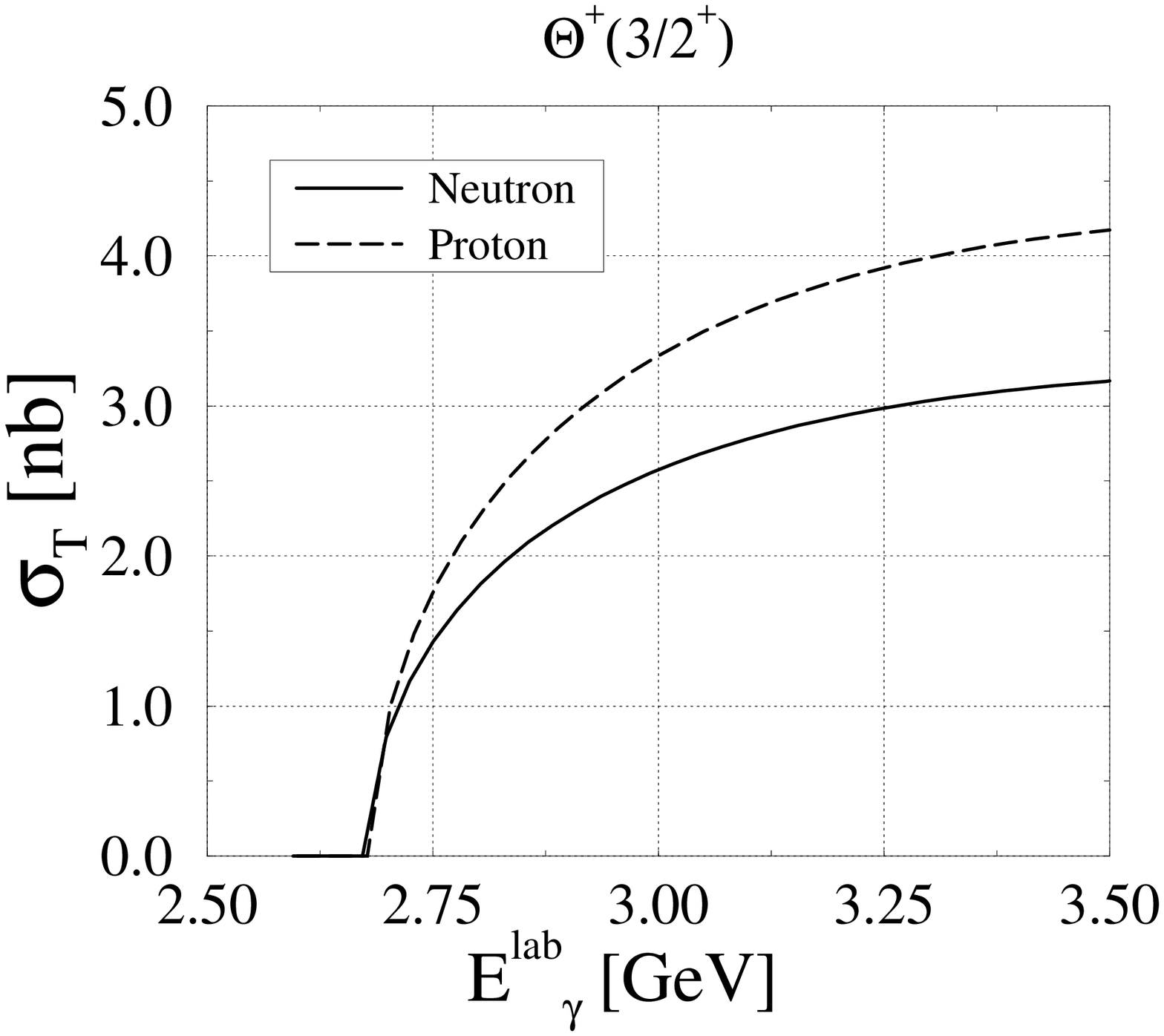}
\includegraphics[width=7cm]{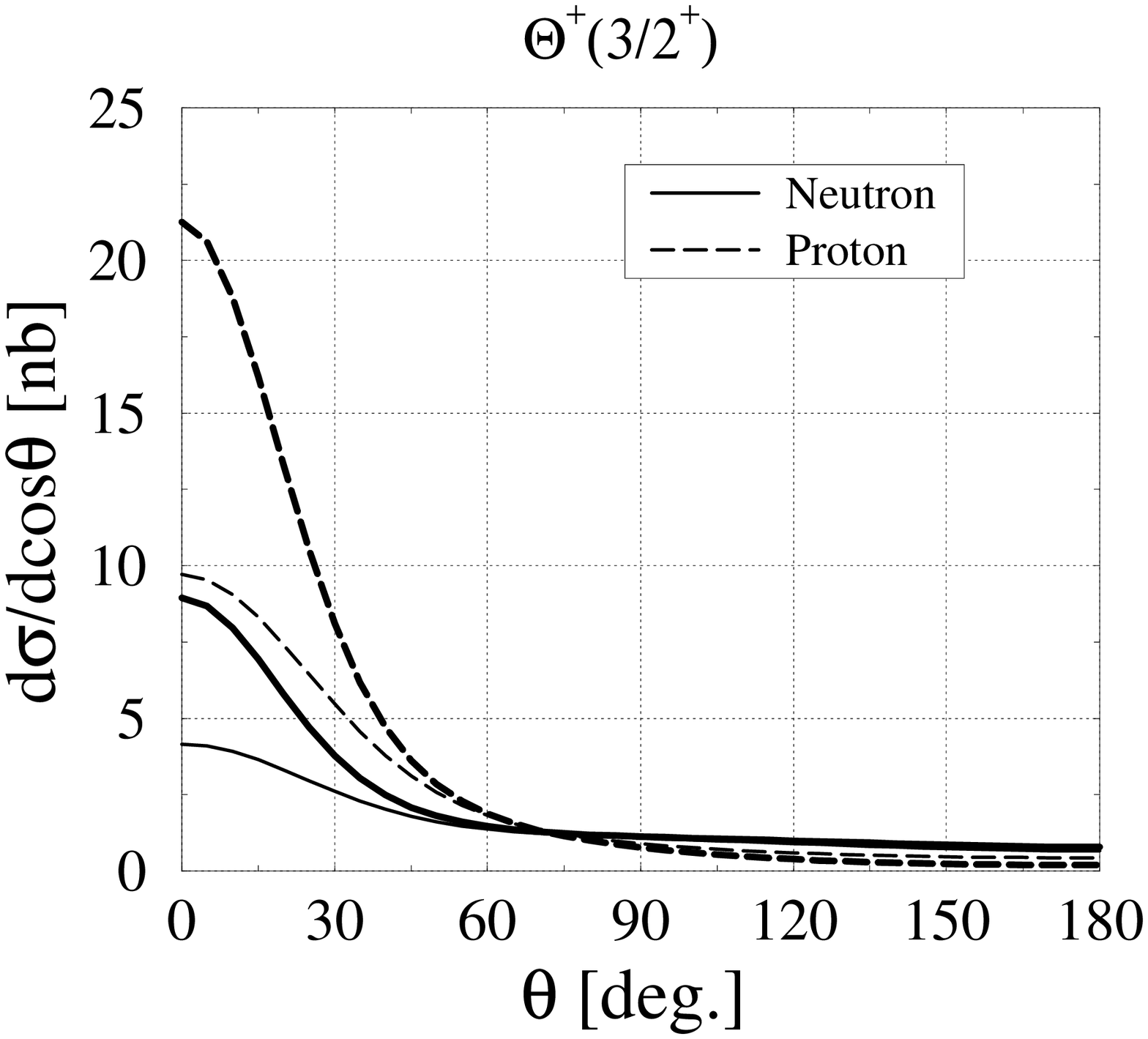}
\end{tabular}
\begin{tabular}{cc}
\includegraphics[width=7cm]{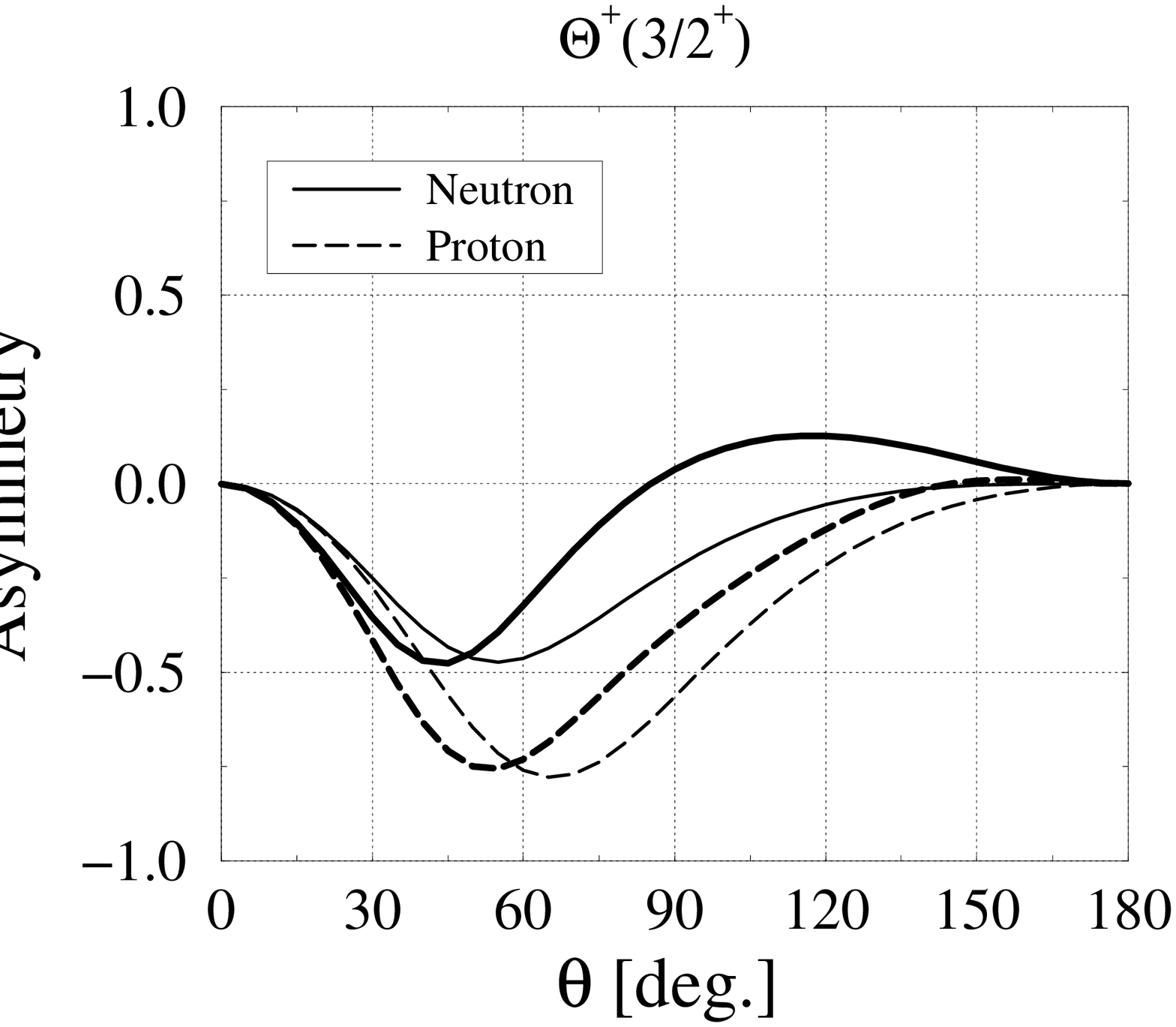}
\includegraphics[width=7cm]{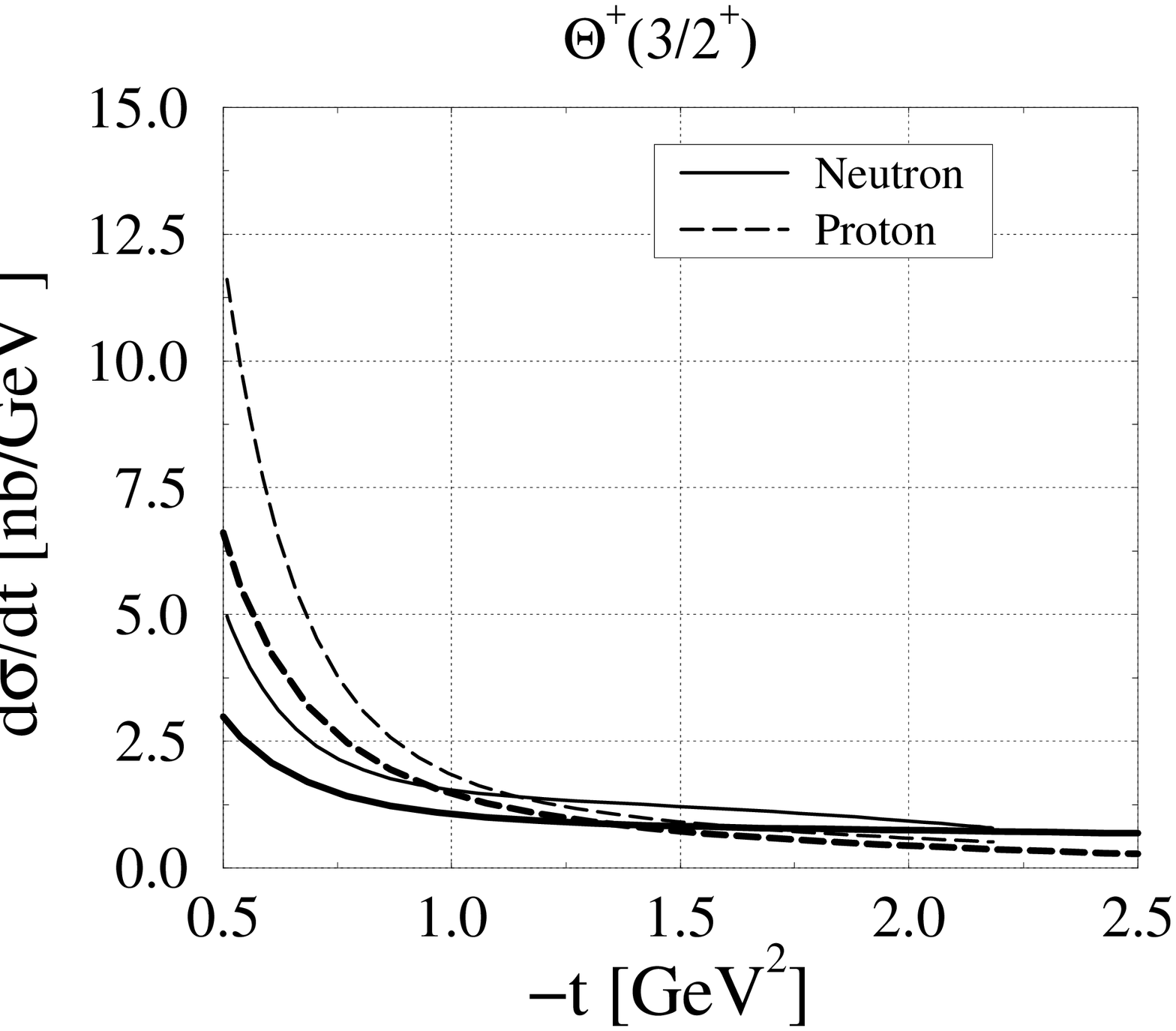}
\end{tabular}
\caption{The total
(upper-left) and differential (upper-right) cross 
sections, the asymmetry (lower-right), and the momentum transfer
$t$-dependence (lower-right) for $\Theta^+(3/2^+)$. The solid and
dashed curves represent the 
results from the neutron and proton targets, respectively.  Thin curves
denote those calculated at $E^{\rm lab}_{\gamma}=3.0$ GeV, while
thick ones stand for those at $E^{\rm lab}_{\gamma}=3.5$ GeV.}     
\label{fig2}
\end{figure}

In Fig.~\ref{fig2} we show the total (upper-left) and differential
(upper-right) cross sections, the asymmetry (lower-left) due to the 
different photon polarizations, and the momentum transfer
$t$-dependence (lower-right) for $\Theta^+(3/2^+)$. The total cross 
sections from the neutron (solid line) and proton (dashed line)
targets are not very much different; the proton case is slightly
larger due to the ratio $g_{\gamma K^0\bar{K}^{*0}}/g_{\gamma
K^+K^{*-}}\sim 1.5$.  The differential cross sections 
are calculated at two 
different photon energies, i.e. $E^{\rm lab}_{\gamma}=3.0$ GeV (thin
curves) and $3.5$ GeV (thick curves).  The angle $\theta$ denotes the one 
between the incident photon and outgoing $K^*$ in the center of
mass frame.  It is clearly shown that the differential cross section
in the forward direction is strongly enhanced; it is mainly due to 
$K$-exchange.  We also find that $\kappa$-exchange 
increases the differential cross section in the forward direction.  The
asymmetry behaves similarly in general for the proton and neutron
targets as shown in the lower-left panel of Fig.~\ref{fig2}. The sign
of the asymmetry is negative when $K$-exchange dominates the process. The
momentum transfer $t$-dependences are drawn in 
the lower-right panel.  The $t$-dependences show again the strong
enhancement in forward scattering. Also, we verified that the
dependence on the coupling constants $g_{\gamma\kappa K^*}$ and
$g_{\kappa N\Theta_{3}}$ is not significant, since the contributionof
$\kappa$ (t(S)) is small as shown in the upper-left panel of
Fig.~\ref{fig1}. Even 
for the case that we use $g_{\gamma\kappa K^*}=2|g_{\gamma KK^*}|$ and
$g_{\kappa N\Theta_{3}}=2|g_{KN\Theta_{3}}|$, only $25\%$ or less difference
appears in the order of magnitudes of the total cross
sections. Furthermore, other observables are not changed much by this
choice.

\begin{figure}[t]
\begin{tabular}{cc}
\includegraphics[width=7cm]{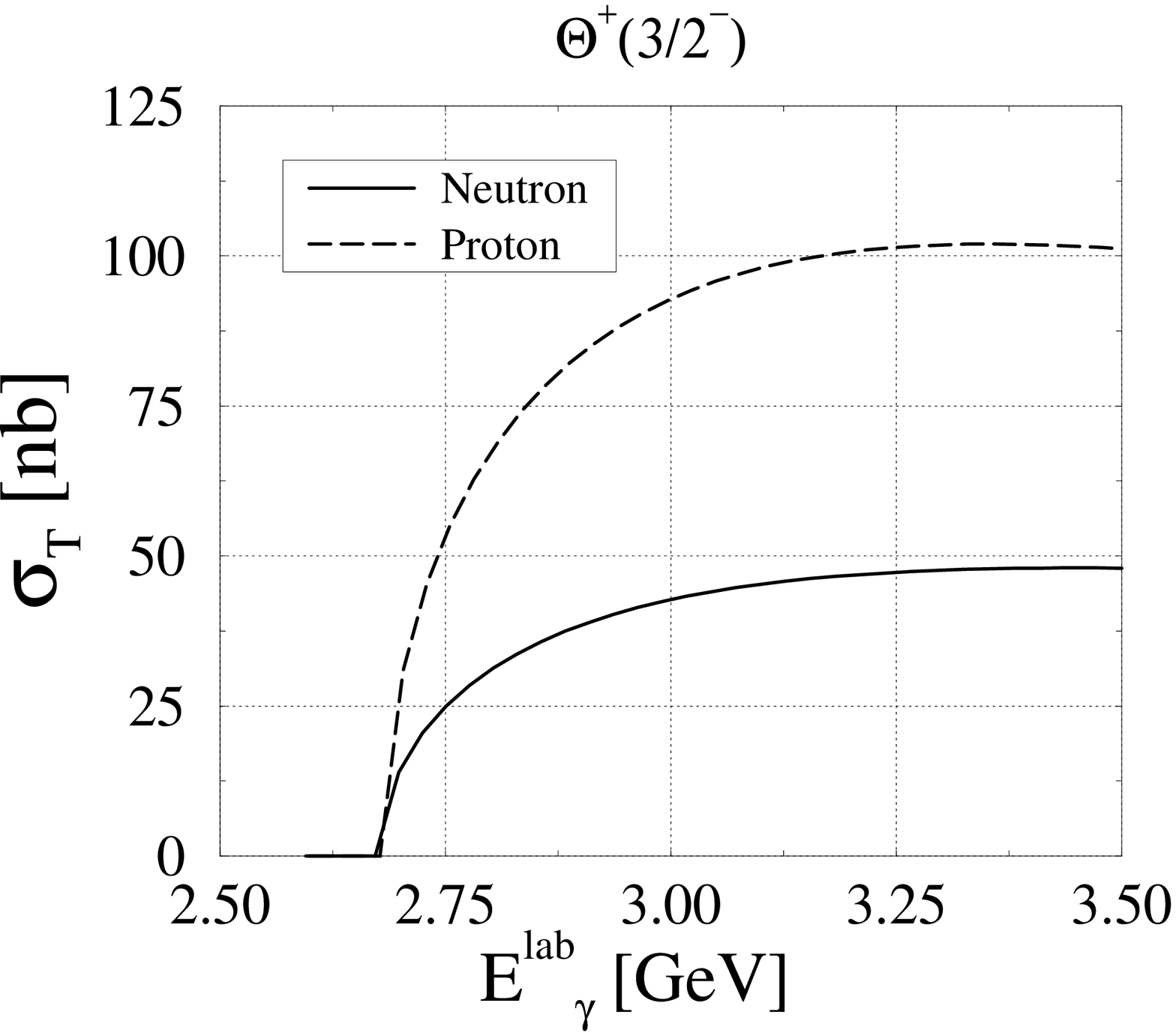}
\includegraphics[width=7cm]{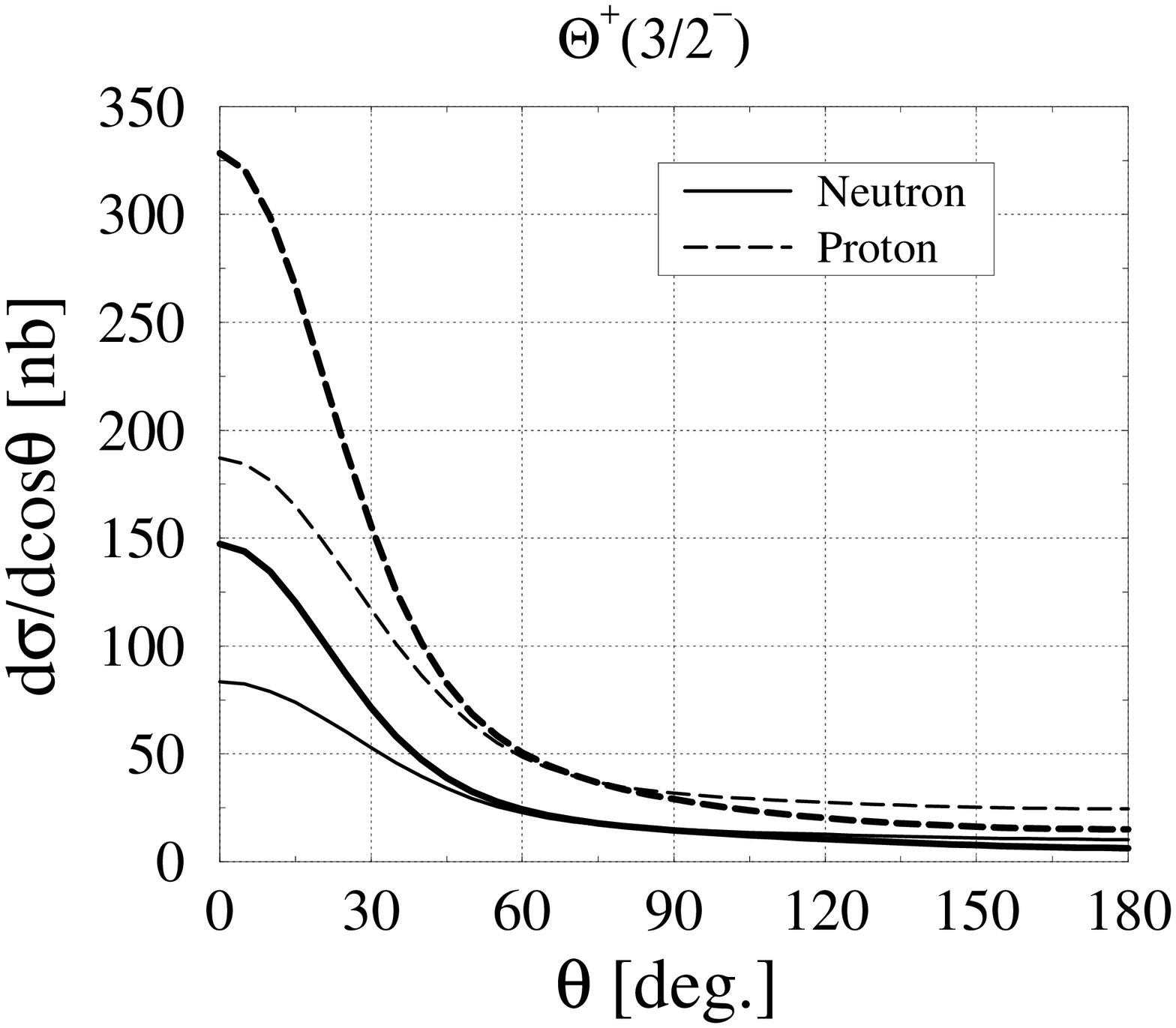}
\end{tabular}
\begin{tabular}{cc}
\includegraphics[width=7cm]{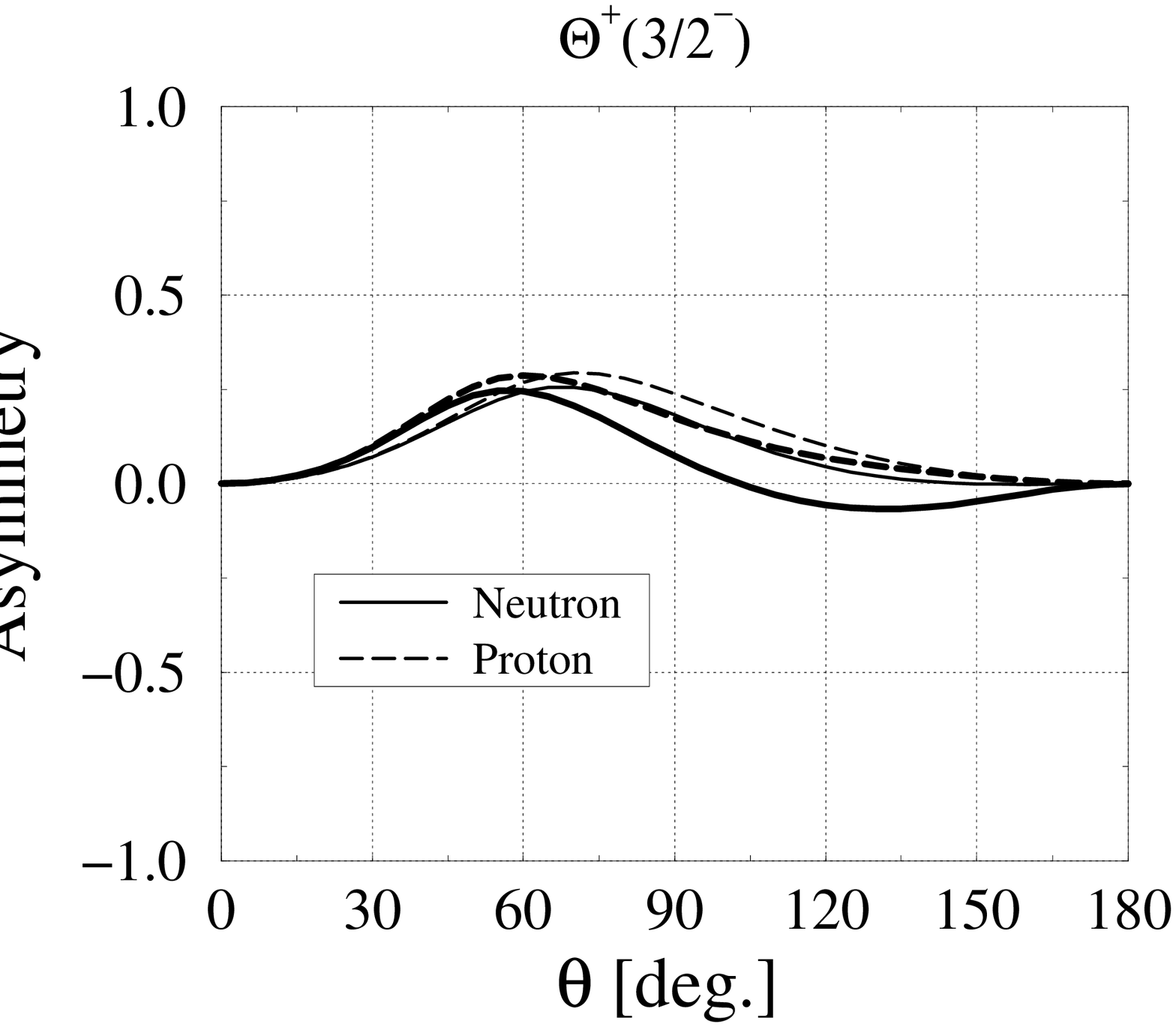}
\includegraphics[width=7cm]{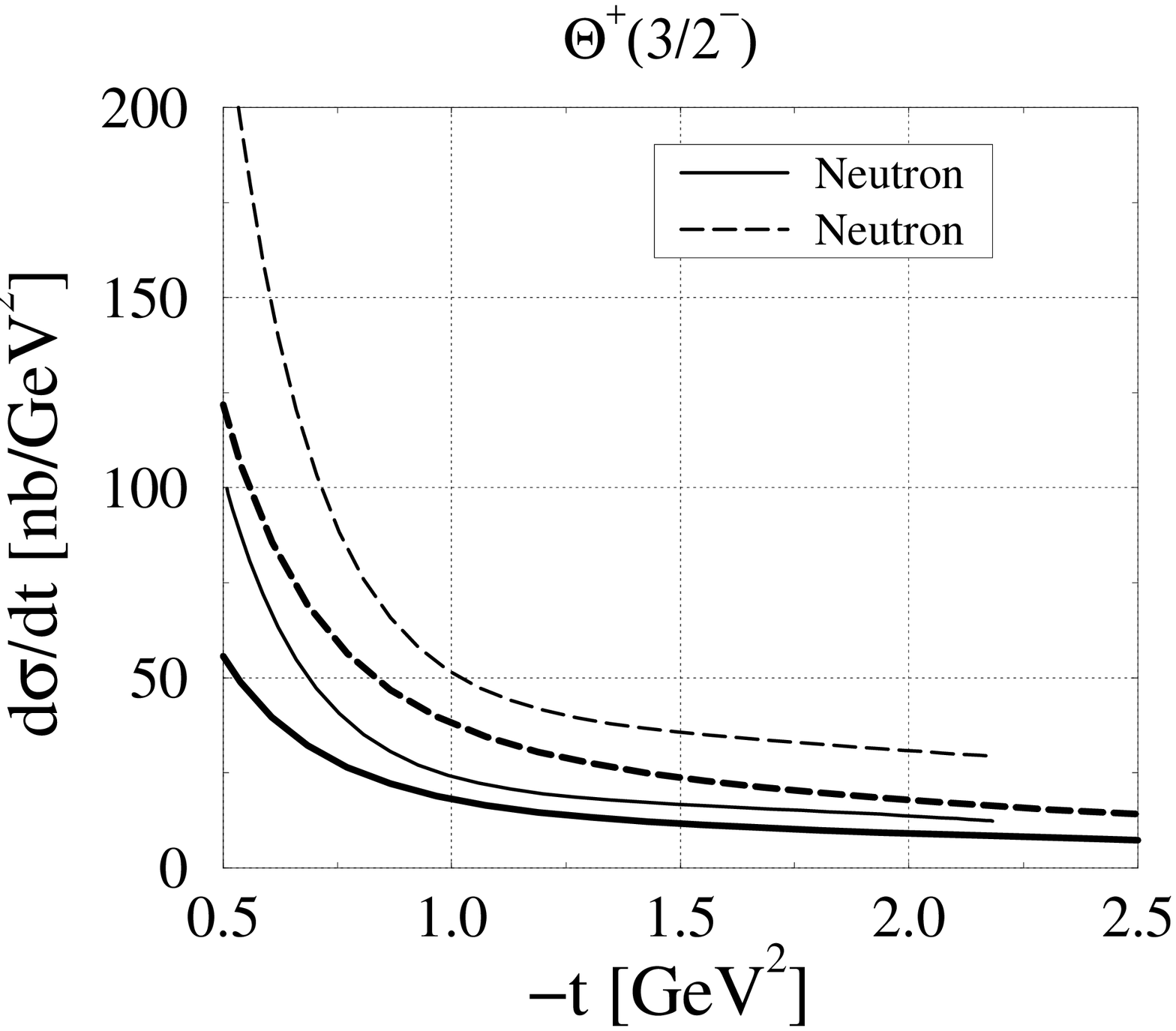}
\end{tabular}
\caption{The total
(upper-left) and differential (upper-right) cross 
sections, the asymmetry (lower-right), and the momentum transfer
$t$-dependence (lower-right) for $\Theta^+(3/2^-)$. The solid and
dashed curves represent the 
results from the neutron and proton targets, respectively.  Thin curves
denote those calculated at $E^{\rm lab}_{\gamma}=3.0$ GeV, while
thick ones stand for those at $E^{\rm lab}_{\gamma}=3.5$ GeV.
}     
\label{fig3}
\end{figure}

Now, we turn to the results for the $\Theta^+(3/2^-)$ depicted in
Fig.~\ref{fig3}.  The total cross sections turn out to be about a few
tens times larger than those for the $\Theta^+(3/2^+)$. The angular
distributions (differential cross sections 
and the momentum transfer $t$-dependence) are rather similar to those
for $\Theta^+(3/2^+)$, since the contributions of $K$- and
$\kappa$-exchanges enhance the forward scattering.  However, the
asymmetries are distinguished clearly from the case of the 
$\Theta^+(3/2^+)$.  The asymmetries for the $\Theta^+(3/2^-)$
production are in general positive when $\kappa$-exchange dominates.
However, if $\kappa$-exchange 
is switched off, the asymmetries becomes similar to those
for the $\Theta^+(3/2^+)$ production with negative sign due to
$K$-exchange dominance, which indicates that 
$\kappa$-exchange plays a key role in distinguishing $\Theta^+(3/2^-)$
from the positive-parity one. 

We note that, however, the dependence on
the couplings of scalar  $\kappa$ in the case of the negative
parity is not ignored, being different from the previous case of
positive parity. This aspect can be easily verified by the curves
shown in the lower-left panel of Fig.~\ref{fig1} in which the
$\kappa$-exchange in $t$-channel, t(S) is the dominant
contribution. Thus, the choice of $g_{\gamma\kappa K^*}=2|g_{\gamma
KK^*}|$ and $g_{\kappa N\Theta_{3}}=2|g_{KN\Theta_{3}}|$ enhances the
magnitudes of the total cross sections by a factor more than
$\sim10$. For instance, we obtain $\sim 420$ nb at
$E_{\gamma}=3.0$ GeV for the $\Theta(3/2^-)$-photoprodution from the
neutron target. Despite the strong dependence on these coupling
constants, the angular distributions are not much affected and show
the strong forward 
enhancement. The asymmetry defined in Eq.~(\ref{asym}) becomes 
all positive for the neutron and proton targets with a 
similar shape as shown in the lower-left panel of
Fig.~\ref{fig2}. 

\begin{figure}[t]
\begin{tabular}{cc}
\includegraphics[width=7cm]{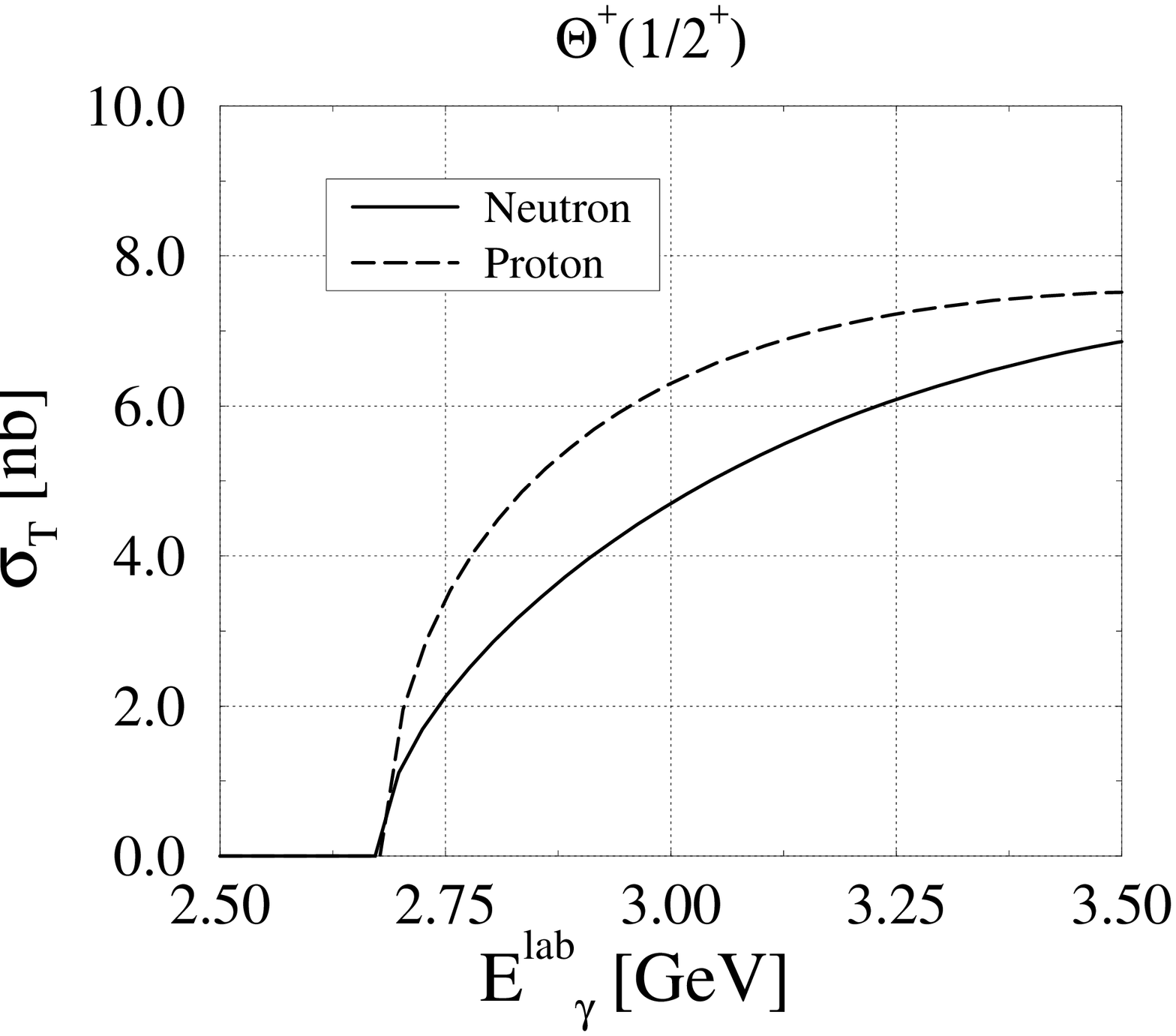}
\includegraphics[width=7cm]{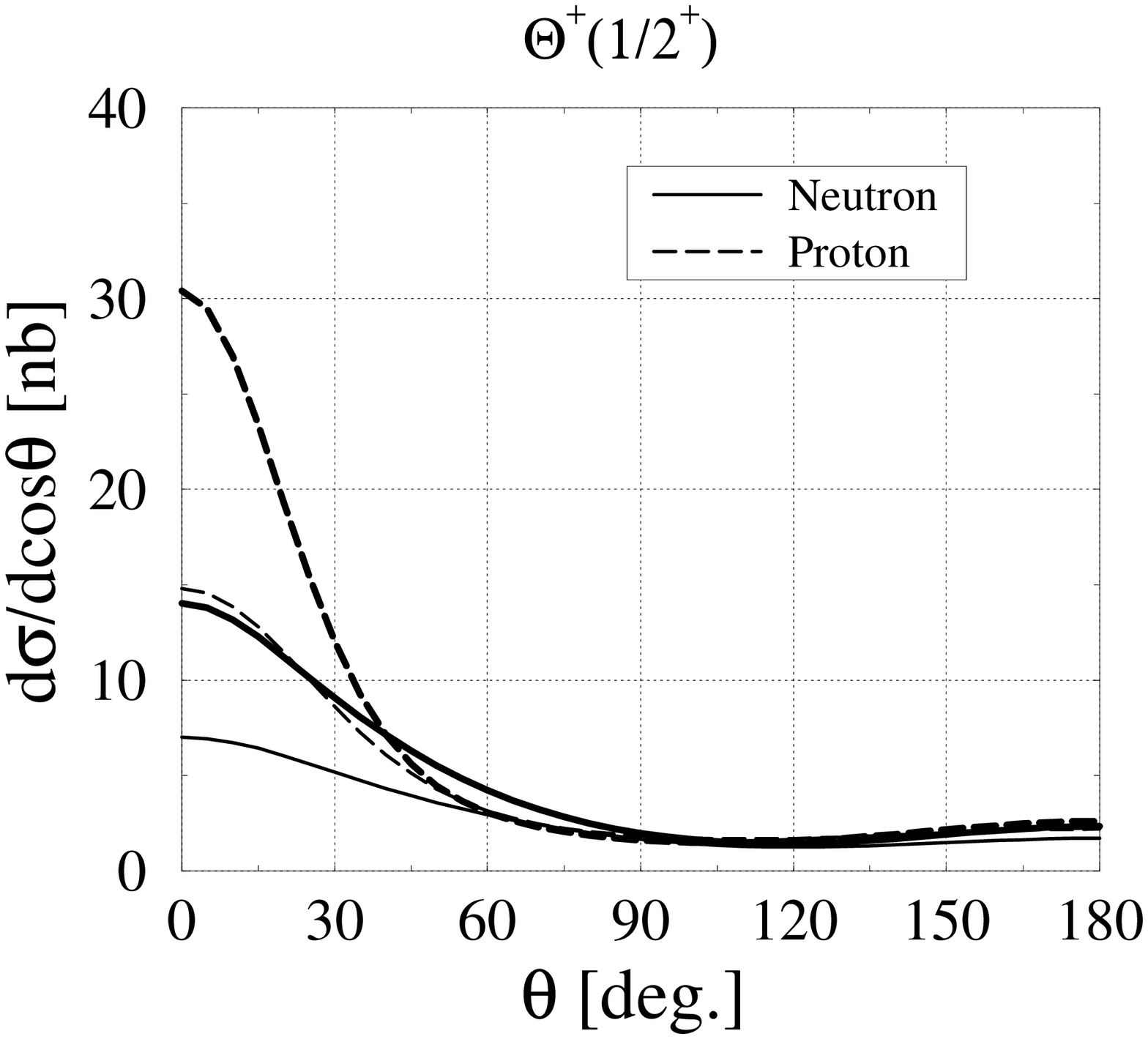}
\end{tabular}
\begin{tabular}{cc}
\includegraphics[width=7cm]{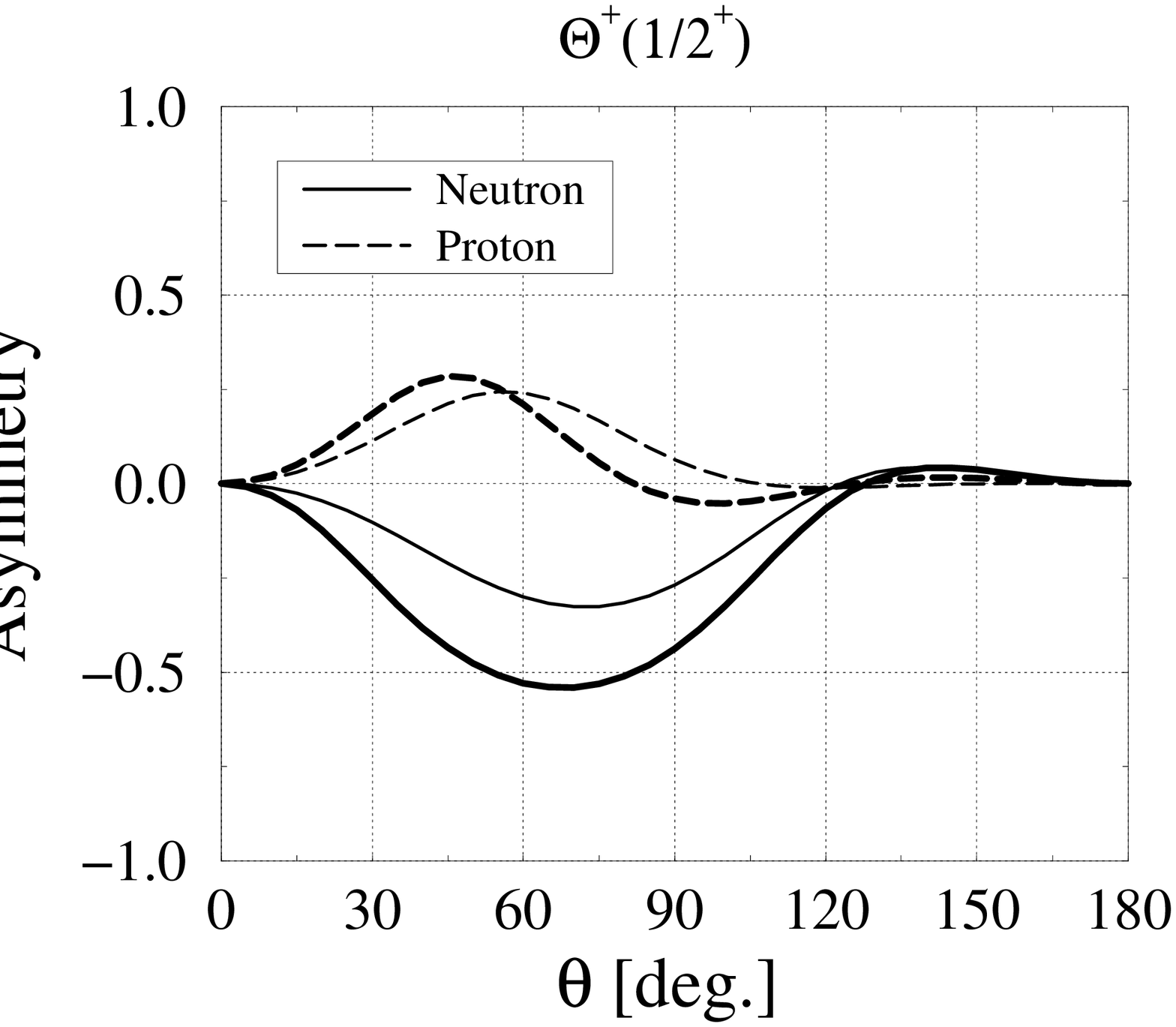}
\includegraphics[width=7cm]{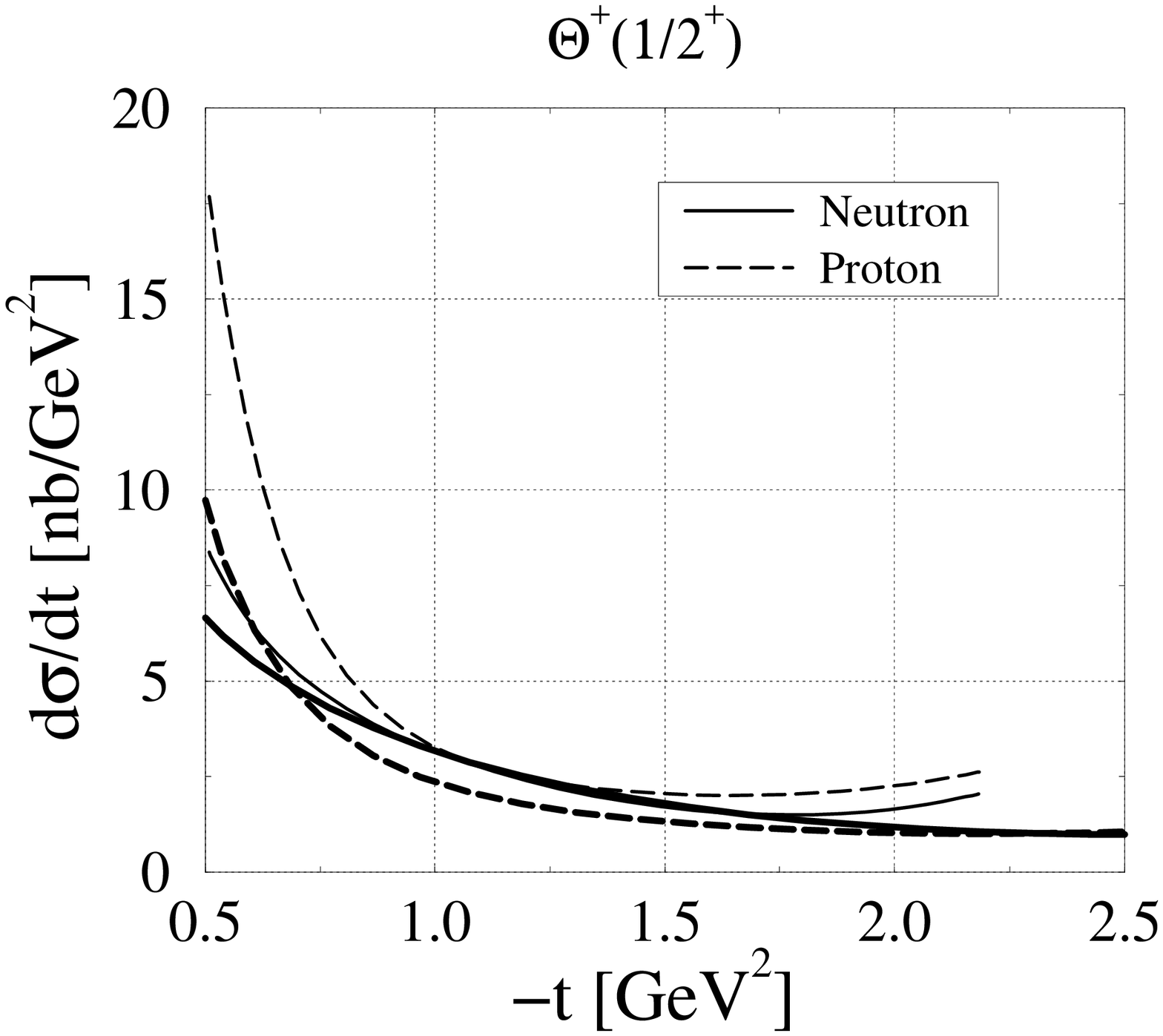}
\end{tabular}
\caption{The total
(upper-left) and differential (upper-right) cross 
sections, the asymmetry (lower-right) and the momentum transfer
$t$-dependence (lower-right) for $\Theta^+(1/2^+)$. The solid and
dashed lines represent the 
results from the neutron and proton targets, respectively. Thin lines
are for the results calculated at $E^{\rm lab}_{\gamma}=3.0$ GeV while
thick lines for done at $E^{\rm lab}_{\gamma}=3.5$ GeV.}     
\label{fig4}
\end{figure}

From here, we compare the results of spin $1/2$ $\Theta^+$ with the
spin $3/2$ $\Theta^+$-photoproduction in Fig.~\ref{fig4}.  Here, we
consider only the case  
of the positive-parity $\Theta^+$, since the cross sections for the
negative-parity one are in general about ten times smaller than those
for the positive-parity $\Theta^+$ (see, for example,
Ref.~\cite{Nam:2003uf}).  However, we note that 
the contribution of $\kappa$-exchange was not considered in the former
studies 
~\cite{Nam:2003uf}.  The total cross sections are of a few
nanobarn order, being similar to and slightly larger than that of
$\Theta^+(3/2^+)$.  We also observe that the angular distribution is
enhanced strongly in the forward direction. The sign of the asymmetry
depends on the type of the target; for the proton target it is positive
while for the neutron one negative. We have checked that the contribution from
the tensor terms proportional to $g^T_{K^*N\Theta_1}$ makes the cross
sections larger only by $\sim10\%$ (see Eq.~(9)) when
$g^T_{K^*N\Theta_1}=|g^V_{K^*N\Theta_1}|$. It also turns 
out that the effects 
from the tensor terms on the angular 
distribution and asymmetry are negligible. However, again, rather 
strong dependence on  
the coupling constants of $g_{\gamma\kappa K^*}$ and $g_{\kappa
N\Theta_{3}}$ are observed as shown in the case of
$\Theta(3/2^-)$. Especially, the asymmetry becomes all positive
having peaks at $\sim70^{\circ}$ for the neutron and proton.

\section{Reaction analysis via the photon and $K^*$ polarizations}
Last but not least, we discuss the analysis of the polarizations of
the photon and the vector $K^*$ meson.  Since the $K^*$ meson can
decay  into the pseudoscalar kaon and pion, it is possible to
determine the polarization state of $K^*$ by the measured azimuthal 
distribution of the kaon and pion. By doing this, we can tell what
meson exchange in the present reaction plays a dominant role.
Similar analysis can be extended to other spin $3/2$ as well as spin $1/2$ 
baryon productions.    

\begin{figure}[t]
\begin{tabular}{cc}
\includegraphics[width=7cm]{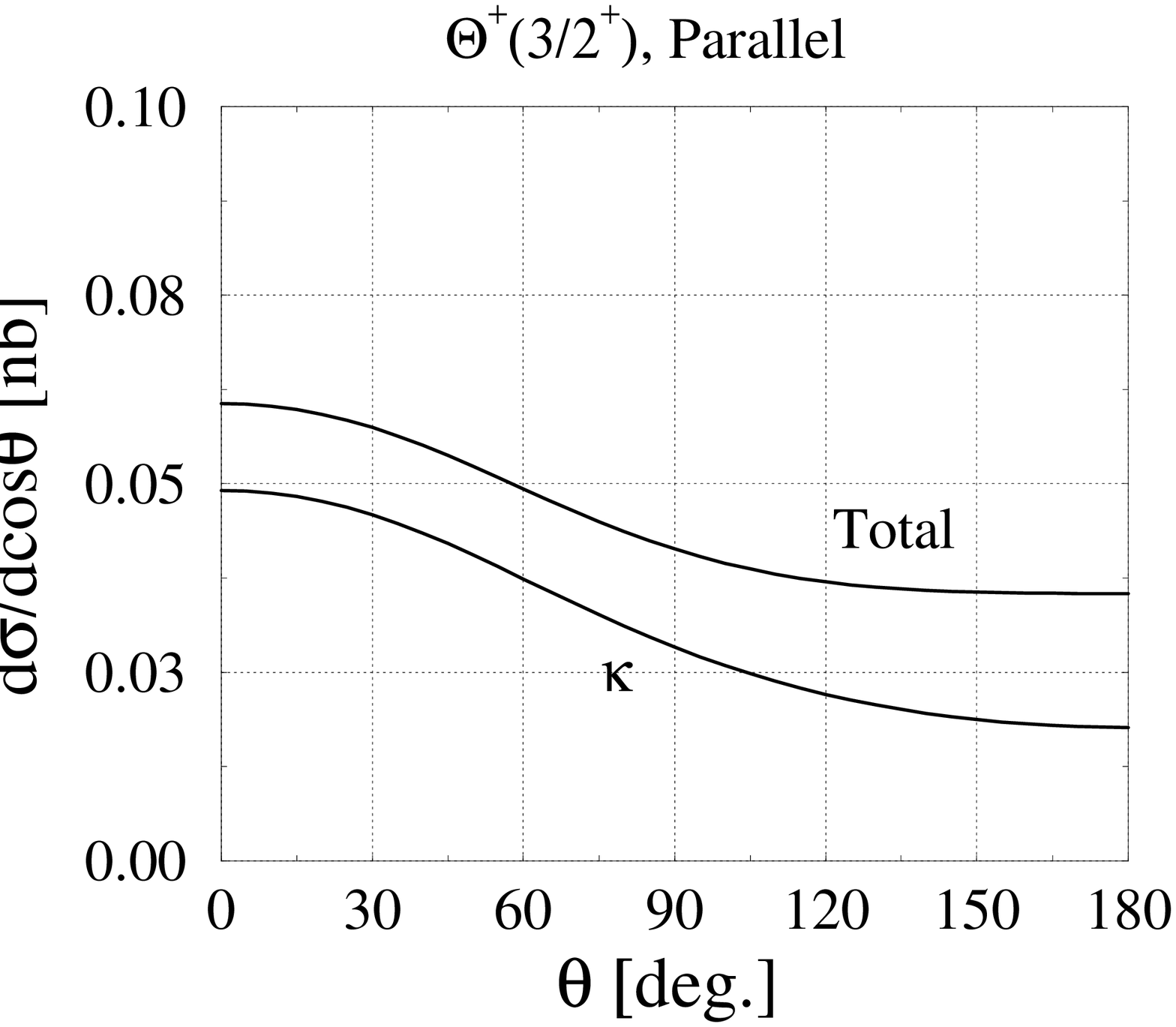}
\includegraphics[width=7cm]{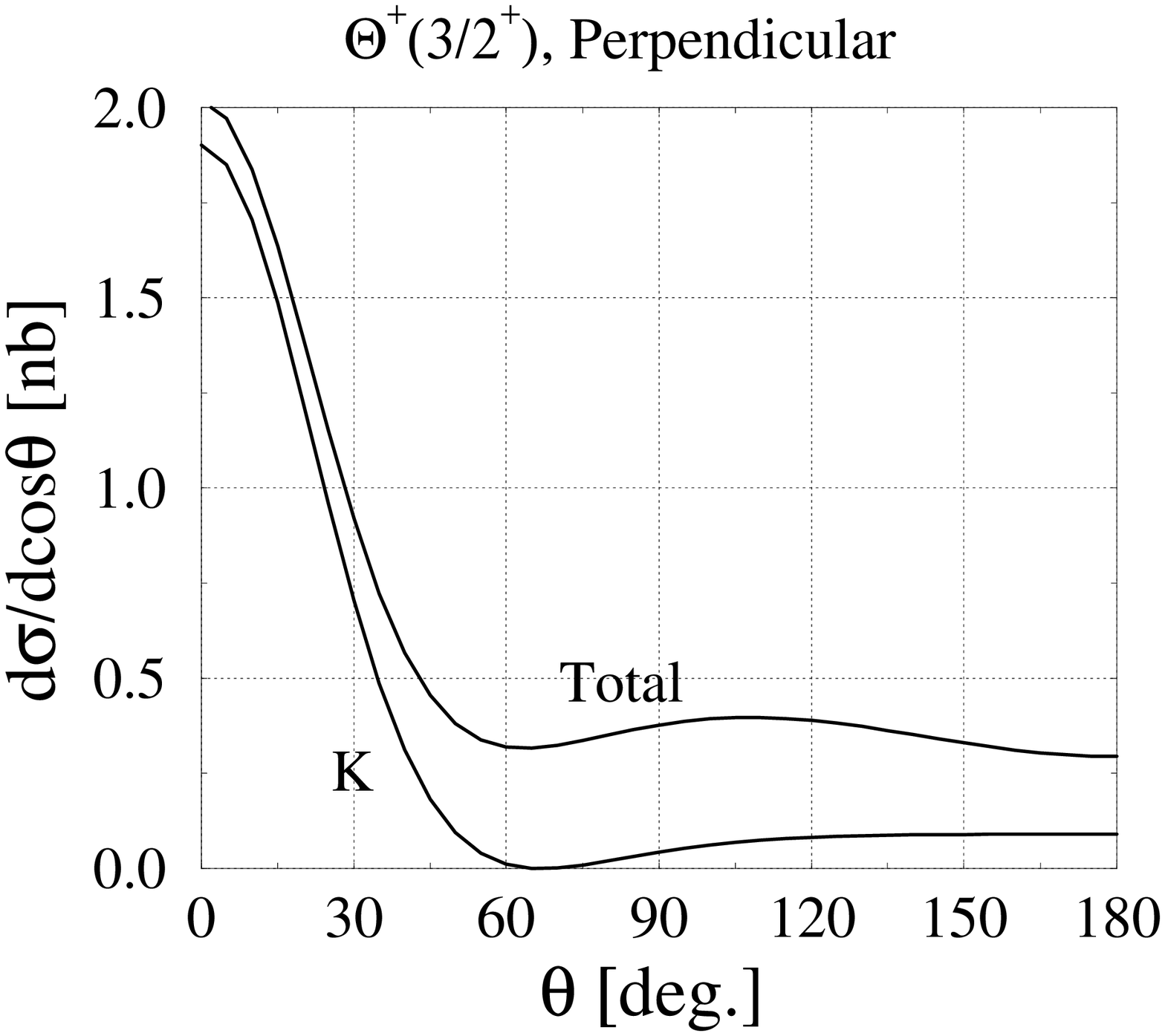}
\end{tabular}
\begin{tabular}{cc}
\includegraphics[width=7cm]{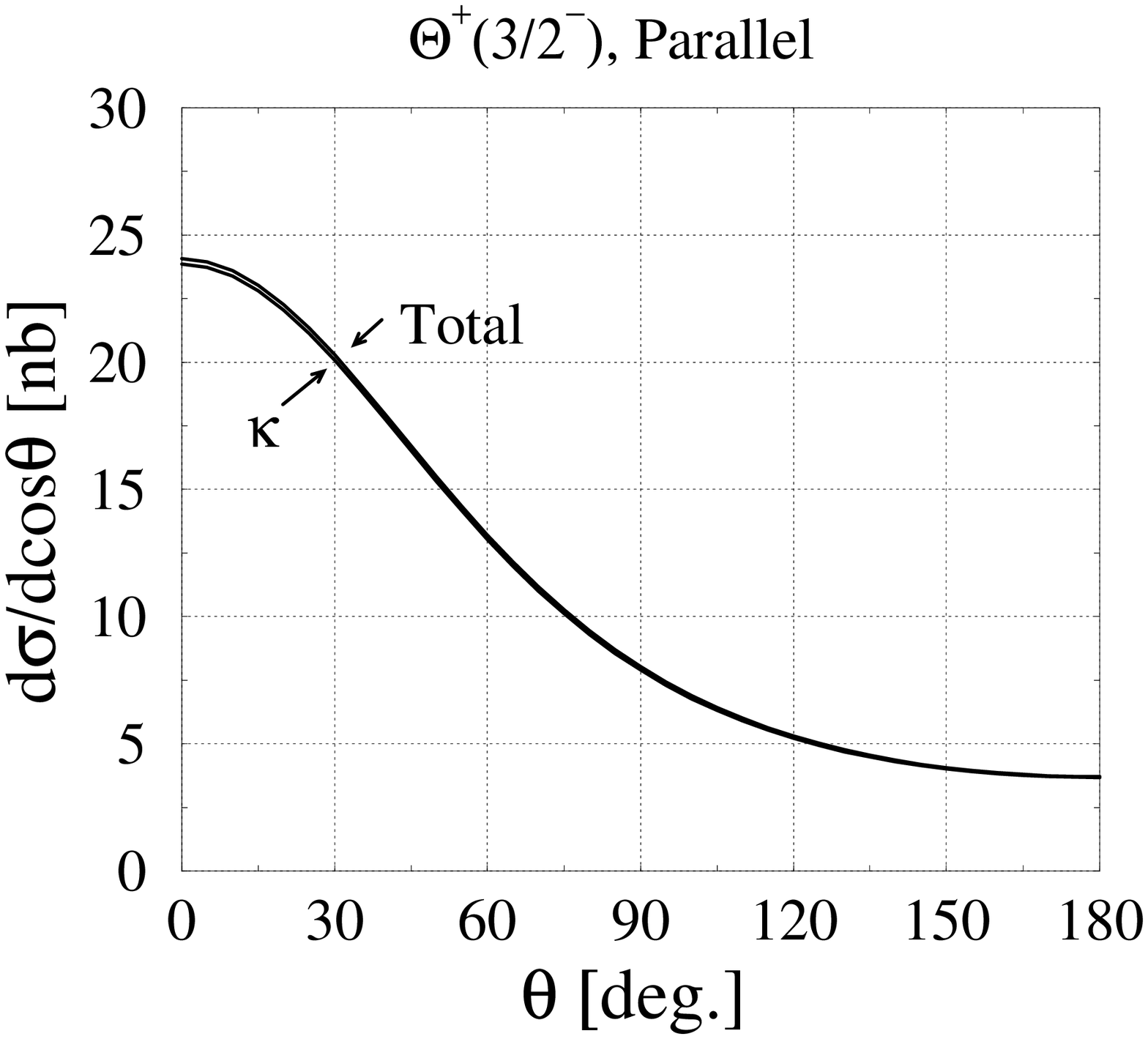}
\includegraphics[width=7cm]{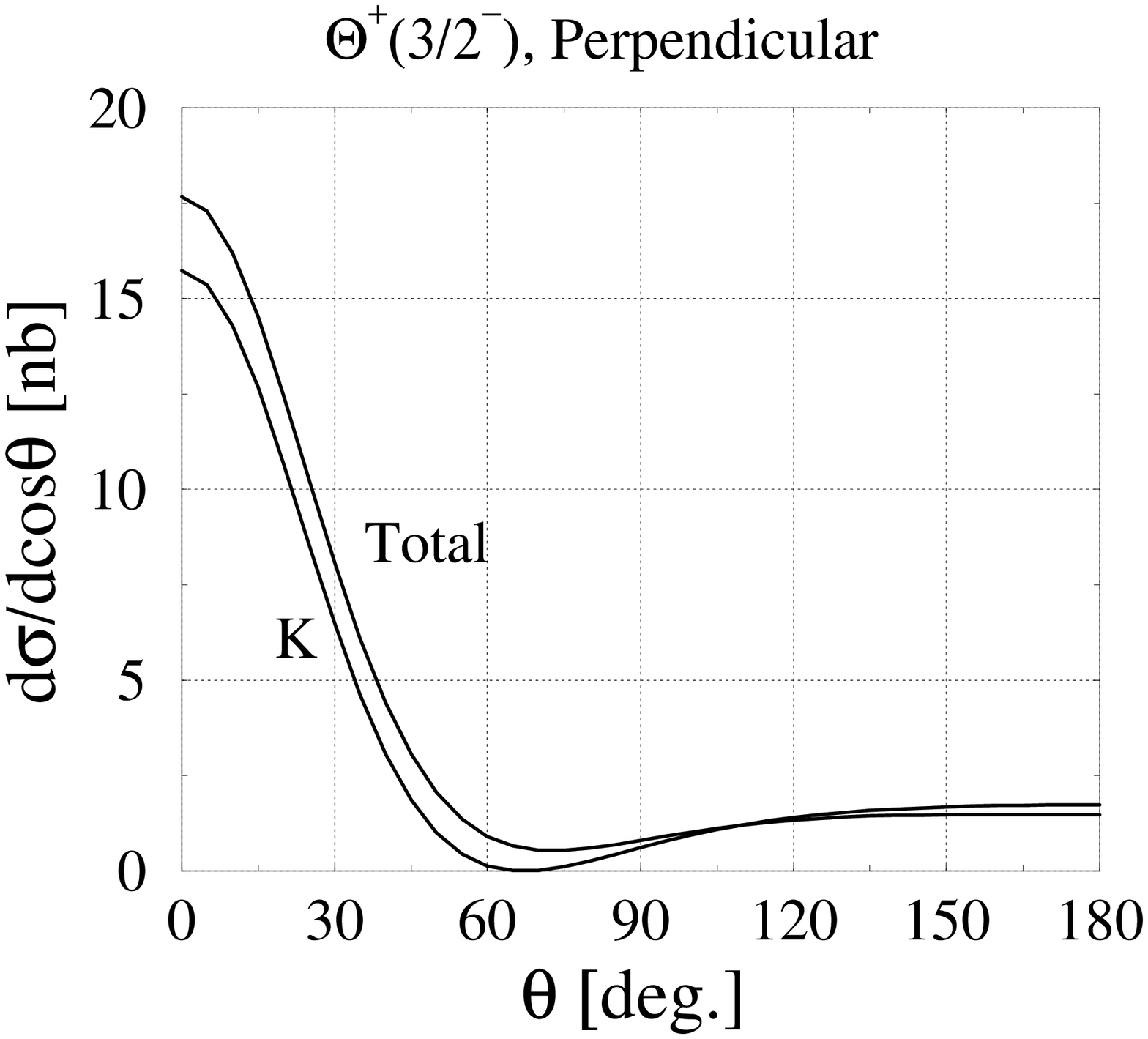}
\end{tabular}
\caption{Differential cross sections when the photon and $K^*$ are
polarized in parallel (left) and perpendicular (right) to each other. We
consider the states of $J^P=3/2^+$ (upper panels) and $3/2^-$ (lower
panels).}      
\label{fig5}
\end{figure}

For this purpose, we first fix the photon polarization to be
perpendicular to the reaction plane. Then, as clearly shown in
Eq.~(\ref{amplitudes1}), the $K^*$-exchange contribution disappears,
since it is proportional to $k_2\cdot\eps_{\gamma}$ in which $k_2$ and
$\eps_{\gamma}$ denote the outgoing $K^*$ momentum and photon
polarization vector, respectively.  Now, let us set the polarization
vector of $K^*$, $\eps_{K^*}$ to be parallel to the direction of  
$\eps_{\gamma}$.  In this case, examining the
$\eps_{\mu\nu\sigma\rho}$ structure of $K$-exchange in
Eq.~(\ref{amplitudes1}), one can easily see that the contribution of
$K$-exchange vanishes.  Thus, as shown in the 
panels on left side of Fig.~\ref{fig5}, only $\kappa$-exchange
survives for both the positive (in the upper panel of
Fig.~\ref{fig5} and negative (in the lower panel of Fig.~\ref{fig5}
parity $\Theta^+$.  We also observe that $\kappa$-exchange dominates
the reaction even when we include all channels, as depicted by the
curve labeled as ``Total'' in Fig.~\ref{fig5}. However, we note that
the strengths of the $\kappa$-exchange contribution depends on the
unknown 
$\kappa N\Theta$ and $\gamma\kappa K^*$ coupling constants.  

We now proceed to examine the case when the two polarization vectors are
perpendicular to each other.  As in the parallel case, the photon
polarization vector is fixed to be perpendicular to the reaction plane
so that $K^*$-exchange can be eliminated.  The corresponding results 
are shown in the right side of Fig.~\ref{fig5}.  The amplitude of
$\kappa$-exchange turns out to be zero, because the term in the bracket
of Eq.~(\ref{amplitudes}) vanishes. Therefore, the contribution
comes only from pseudoscalar $K$-exchange.  Experimentally, the
comparison of the two polarization combinations,
$\eps_{\gamma}\perp\eps_{K^*}$ and $\eps_{\gamma}\parallel\eps_{K^*}$,
provide information of the strengths of the $KN\Theta$ and $\kappa
N\Theta$ coupling constants. 

The bump or the increase in the differential cross sections for
$\theta\gsim 60^{\circ}$ as shown in the right
side of Fig.~\ref{fig5} is mainly due to the contact term
contribution.  The total contributions do not differ much from the
cases with the $K$-exchange contribution only.  Interestingly, the
results for the two different parities of $\Theta^+$ are rather
similar each other except for the order of magnitudes, since the
polarization dependence arises only from the structure of the $\gamma
K^*M(K,K^*,\kappa)$ coupling, but not from that of $MN\Theta^+$ one,
which carries the information of the parity of $\Theta^+$. 
 
The polarization analysis of the photon and vector $K^*$ sheds light on
determining which meson exchange is dominant in the present
reaction.  Though we do not show the results for the
$\Theta^+(1/2^+)$-photoprodcution explicitly here, we verified that
the similar 
conclusion was drawn.  We notice that this analysis may also be of
great use in determining which meson is the most prominent in
general $\gamma N\to M(1^-)B$ reactions, since the  
method discussed here is based only on the structure of the
photon-meson-meson vertices, but not of vertices including baryons. 

\section{Summary and Conclusion}
We have investigated the photoproduction of the exotic pentaquark
baryon $\Theta^+$ via the reaction process $\gamma N\to
\bar{K}^*\Theta^+$, assuming that $\Theta^+$ has spin $3/2$.  The
effective Lagrangian approach was employed with phenomenological form
factors~\cite{Nam:2005jz,Nam:2005uq}.  We used the coupling
constant for the $K^*N\Theta(3/2)$ vertex estimated from 
the constituent quark model.  We also considered scalar meson
$\kappa(800,0^+)$-exchange.  We assumed
the following relations for the coupling constants; $g_{\gamma\kappa K^*}=g_{\gamma KK^*}$ and $g_{\kappa N\Theta}=g_{KN\Theta}$ as a
trial. The main results of the present work are summarized in
Table.~\ref{table2}.       

In the present work, we did not find large difference
between the total cross sections from the neutron and proton targets,
which is different from the conclusion of the previous work of $\gamma 
N\to \bar{K}\Theta^+(3/2)$~\cite{Nam:2005jz}.  The reason lies in the
fact that the contact term in the present case does not provide a
large contribution to the cross sections, compared to other
meson-exchange.  These differences between the
$\Theta^+$-photoproductions with the pseudoscalar $K$ and with the
vector $K^*$ can be useful to determine the spin quantum number of the
$\Theta^+$ baryon.  We estimated the total cross 
sections for the present reaction qualitatively as follows:
$\sigma_{3/2^+}\sim 1.5$ nb and $\sigma_{3/2^-}\sim 50$ 
nb for the energy regions of $E_{\rm th}\lsim E^{\rm lab}_{\gamma}\lsim
3.5$ GeV for both the neutron and proton targets.  We notice that
there is the model dependence due to the coupling constants of
$\kappa$-exchange, in particular, in the case of $\Theta^+(3/2^-)$.
However, the tendency
$\sigma_{\Theta^+(3/2^+)}<\sigma_{\Theta^+(3/2^-)}$ is rather stable, 
since psuedoscalar $K$-exchange which has a less dependence on the
model parameters is the most
dominant contribution in the present reaction.  

In angular distributions, we observed a large enhancement in the
forward region due to the $t$-channel dominance ($K$- and
$\kappa$-exchanges) for both the 
spin $1/2$ and spin $3/2$ cases.  From these
observations, we expect that in the laboratory frame, there must be
even stronger forward enhancement for the outgoing $K^*$.   The asymmetry
shows relatively clear difference between the positive and negative
parities of the $\Theta^{+}(3/2) $, though there is one caveat: once
we know the strengths of the coupling constants 
$g_{\gamma\kappa K^*}$ and $g_{\kappa N\Theta}$. We also compared the
present results to those from the reaction with the $\Theta^+(1/2^+)$.

Finally, an analysis was proposed to determine which meson exchange is
dominant in the $t$-channel, with the photon and $K^*$ polarizations being  
explicitly considered.  It was observed that scalar meson
$\kappa$-exchange only survives when the polarizations of the photon
and $K^*$ are parallel.  On the contrary, when these
polarizations are perpendicular to each other, pseudoscalar
$K$-exchange turns out to be dominant.  This analysis may be applied to
a general reaction $\gamma N \to M(1^-)B$.  

We note that the coupling constants $g_{\gamma \kappa K^*}$ and $g_{\kappa
N\Theta}$, being important in the present investigation, are not known
well. Especially, the asymmetry is affected much by the different
choices of the coupling constants for the cases of $\Theta(3/2^-)$ and
$\Theta(1/2^+)$ whereas the cross sections are changed only in the order of
the magnitudes. 
Considering the rather small values shown in Table.~\ref{table2}, it might be
rather difficult to observe a clear peak from the present reaction process in
the present experimental facilities. However, since we once again observed
strong forward scattering enhancement which could be measured most
appropriately by LEPS, it is expected that different experimental setup may
obtain sizable statistics for the indication of $\Theta^+$ for the present
reaction process. 
      
\begin{table}[t]
\begin{tabular}{|c|c|c|c|c|c|c|}
\hline
$J^P$&\multicolumn{2}{c|}{$3/2^+$}&
\multicolumn{2}{c|}{$3/2^-$}&
\multicolumn{2}{c|}{$1/2^+$}\\
\hline
Target&$n$&$p$&$n$&$p$&$n$&$p$\\
\hline
$\sigma$ at $E_{\gamma}^{\rm lab}=3.0$ GeV&$\sim2.5$ nb&$\sim3.2$
nb&$\sim 40$ nb&$\sim90$ nb&$\sim4$ nb& $\sim5.5$ nb\\ 
\hline
\end{tabular}
\caption{Main results of the $\Theta^+$-photoproduction via $\gamma N\to
\bar{K}^*\Theta^+$.}
\label{table2}
\end{table}

\section*{Acknowledgment}
We are very grateful to J.~K.~Ahn, V.~Koubarovski, T.~Nakano, and
T.~Hyodo for fruitful discussions.  The work of S.I.N. has been
supported in part by the scholarship from the Ministry of Education,
Culture, Science and Technology of Japan.  The work of A.H. is 
supported in part by the Grant for Scientific Research ((C)
No.16540252) from the Education, Culture, Science and Technology of
Japan. The work of H.C.K. and S.I.N. was supported by the Korea
Research Foundation Grant funded by the Korean Government (MOEHRD)
(KRF-2005-202-C00102).   
 
\end{document}